\documentstyle[aps]{revtex}
\newtheorem{sn}{}
\renewcommand{\thesn}{%
               \arabic{section}}

\newtheorem{df}{~~~{\sl Definition}}[sn]
\newtheorem{th}[df]{~~~Theorem}
\newtheorem{pp}[df]{~~~{\sl Proposition}}
\newtheorem{lm}[df]{~~~{\sl Lemma}}
\newtheorem{co}[df]{~~~{\sl Corollary}}
\newtheorem{ex}{~~~{\sl Example}}

\newcommand{\qed}{$\Box$}

\begin{document}

\title{A generalized Weyl relation approach to 
the time operator \\ and its connection 
to the survival probability 
\thanks{to appear in Journal of Mathematical Physics } 
}

\author{Manabu Miyamoto 
\thanks{E-mail: miyamo@hep.phys.waseda.ac.jp } 
\\
\sl Department of Physics, Waseda University, Tokyo 169-8555, 
 Japan}
\date{\today}

\maketitle

\begin{abstract}
The time operator, 
an operator which satisfies the canonical commutation relation with 
the Hamiltonian, %in quantum systems 
is investigated, 
%on the basis of the axiomatic quantum mechanics. 
%As a basis of this investigation,  
%the analysis of the time operator, 
on the basis of 
a certain algebraic relation 
for a pair of operators $T$ and $H$, 
%is proposed, 
where $T$ is symmetric and $H$ self-adjoint.  
%which is 
%as generators of unitary group, is proposed.   
%When  $T$ is  self-adjoint, 
%this relation coincides the Weyl relation.  
%similar to the Weyl relation, 
%
%where $T$ is 
%a symmetric operator and $H$ a self-adjoint 
%operator on a Hilbert space, 
%corresponding to the time operator and the Hamiltonian respectively. 
%A difference between these relations is that 
%$T$ being symmetric is allowed, 
%whereas $H$ remains self-adjoint, 
%for the former 
%
This relation is equivalent to the Weyl relation, 
in the case of self-adjoint $T$, 
and 
is satisfied by the Aharonov-Bohm time operator $T_0$ 
and the free Hamiltonian $H_0$ 
for the one-dimensional free-particle system. %, 
%by regarding $T_0$ and $H_0$ respectively as $T$ and $H$. 
%$T_0$ is the Aharonov-Bohm 
%time operator for this system. 
In order to see  
the qualitative properties of $T_0$, %and $H_0$, 
%such independent of the explicit form of this operator, 
%
%$T_0$ and $H_0$, 
%Thus the consideration of this algebraic relation 
%is equivalent to it of 
%the qualitative properties of $T_0$ and $H_0$, 
%independent of the explicit forms of $T_0$ and $H_0$. 
%supposing such operators $T$ and $H$ exist, 
%
%the general properties of 
the operators $T$ and $H$ satisfying 
this algebraic relation 
%such as the spectrum and the uncertainty relation 
are examined. 
In particular, it is shown that 
the standard deviation of 
$T$ is directly connected to the survival probability, 
and $H$ is absolutely continuous. 
%From these facts, 
Hence, 
it is concluded that 
the existence of the operator $T$ implies 
the existence of scattering states.  
It is also shown that the minimum uncertainty states 
do not exist. 
Other examples of these operators $T$ and $H$, 
%such that $H$ is considered as the Hamiltonian 
%for another system, 
than the one-dimensional free-particle system, 
%for which there are time operators, 
are demonstrated. 
%briefly discussed 
%by using results of the above mentioned and the theory of 
%Schr{\"o}dinger operators.

\vspace{5mm}

\noindent 
PACS numbers:\ 03.65.-w,\ 03.65.Db,\ 02.30.Sa

\end{abstract}

%\newpage

%\begin{flushleft}

\section{Introduction} \label{sec:1}

%\end{flushleft}

The concept of the time operator is strongly connected with 
the time-energy uncertainty relation. 
The time operator, denoted by  $T$, is usually defined to satisfy 
the canonical commutation relation (CCR) with 
the Hamiltonian $H$: $[T, H]=i$~
(see \cite{ko} and the references therein). 
If such an operator were defined consistently 
on the Hilbert space corresponding to a 
certain quantum system, 
then the time-energy uncertainty relation 
could be automatically reduced from the Cauchy-Schwarz inequality, 
as in the case between the position and momentum operators 
on $L ^2 ({\bf R}^1)$. 
For instance, if we take the operator $T_0$ 
suggested by Aharonov and Bohm \cite{aha}, 
as a time operator 
for the one-dimensional free-particle system (1DFPS), 
we formally have $[T_0, H_0]=i$ and derive the uncertainty 
relation between $T_0$ and $H_0$. Here 
$H_0 := P^2 /2$ is the free Hamiltonian for the 1DFPS, and 
$T_0$ is defined as 
\begin{equation}
T_0 := \frac{1}{2} \left( Q P^{-1} + P^{-1} Q \right),  
 \label{eqn:1.1}
\end{equation}
where $Q$ and $P$ are the position and momentum operators 
on $L ^2 ({\bf R}^1)$ (more precise definition is given 
in Sec.\ \ref{sec:3}). 
$T_0$ is often called the Aharonov-Bohm time operator. 
It is, however, not clear whether the inverse $P^{-1}$ could be well-defined. 
We should also remember the criticism posed by Pauli \cite{pa}, 
although it is not rigorous, that 
the time operator can not necessarily be defined for all quantum systems 
without contradiction. 
Furthermore the physical meaning of the time operator, if any, still 
remains unclear. 

We shall base our discussion on the axiomatic quantum mechanics. 
Then it is possible to comment on the above difficulties 
from the axiomatic points of view. 
We first see that the inverse $P^{-1}$ is a well-defined 
self-adjoint operator on $L^2 ({\bf R}^1)$ (more details 
are given in Sec.\ \ref{sec:3}). 
Recently, the operator $T_0$ was shown to be well-defined, 
and its mathematical character was clarified, 
through the study of the time-of-arrival problem \cite{eg}. 
Observe that in Pauli's criticism, 
it is implicitly assumed that 
if there exists a self-adjoint operator $T$ which satisfies the CCR 
with the Hamiltonian $H$ for some system, 
\begin{equation}
  TH \psi - HT \psi = i \psi ,~~ 
  \forall \psi \in \mbox{\rm Dom}(TH)\cap \mbox{\rm Dom}(HT), 
\label{eqn:1.2}
\end{equation}
%$\forall \psi \in \mbox{\rm Dom}(TH)\cap \mbox{\rm Dom}(HT)$, 
one would be able to derive the following relation 
\begin{equation}
 He^{i\epsilon T} \psi = e^{i\epsilon T} (H+\epsilon ) \psi ,~~
  \forall \psi \in \mbox{\rm Dom}(H),~
  \forall \epsilon \in {\bf R}^1 .
\label{eqn:1.3} 
\end{equation} 
%$\forall \psi \in \mbox{\rm Dom}(H)$ and 
%$\forall \epsilon \in {\bf R}^1 $. 
We have to be careful, however, 
about this kind of logic, since it is not generally true 
whereas its converse is true.  
For example, consider a pair of operators, 
the position and momentum operators 
on $L^2 ([0,1])$, $Q$ and $P$, 
for the above $T$ and $H$, respectively \cite{ak}. 
They satisfy Eq.\ (\ref{eqn:1.2}) 
but Eq.\ (\ref{eqn:1.3}) is satisfied only for 
the particular values $\epsilon = 2\pi n, ~n \in {\bf Z}$, 
since, in order that $P$ be self-adjoint, 
the domain $\mbox{\rm Dom}(P)$ has to be supplemented with 
a boundary condition $\psi (0) = \theta 
\psi (1)$ with a fixed $\theta \in {\bf C},\ |\theta |=1 $, 
$\forall \psi \in \mbox{\rm Dom}(P)$ \cite{ga}. 
Furthermore, 
%we usually consider that observables are represented by 
%self-adjoint operators, however 
there is no a priori reason why we 
have to consider the time operator an observable, that is, 
a self-adjoint operator: 
we don't have any interpretation of the 
time operator as an observable. In this paper, we shall 
require the time operator be symmetric, 
satisfying Eq.\ (\ref{eqn:1.2}) 
with the Hamiltonian, but not necessarily be self-adjoint. 

The investigation of the time operator is important to 
understanding the time-energy uncertainty relation,  
and may have a significance for the analysis of the dynamics 
of quantum systems. 
A reason for the latter is that 
the time operator is directly connected to the Hamiltonian 
through the CCR, and this is algebraically so strong relation 
between operators as to prescribe qualitative aspects of 
their spectra, we can expect that the time operator 
brings us information about qualitative aspects of 
the time evolution of quantum systems. 
Hence, our purposes here are 
to examine for which quantum 
systems such a symmetric time operator is allowed to exist 
consistently, and to disclose its relevance 
to the dynamics 
of the quantum system under consideration. 
%, and to 
%clarify the physical meaning of the time operator 
%by classifying such systems. 

From what is mentioned above, the investigation of the time 
operator is involved in that of the commutator 
(not necessarily canonical). 
The connection between the commutator of the form 
$[H , iA] =C$ ($C \geq 0$) 
and the spectra of self-adjoint operators, 
$H, A$ and $C$, has been 
widely studied by 
Putnam \cite{pu}, Kato \cite{ka2}, Lavine \cite{la}, \cite{la2},  
and others (see also \cite{re3}). 
We here, however, restrict our consideration to 
the more strong form,  
which will be called the ``$T$-weak'' Weyl relation.

%We first introduce a notion of the `$T$-weak' 
%Weyl relation between a pair of operators on the Hilbert space. 
%In a certain sense this relation 
%is the converse of Eq.\ (\ref{eqn:1.3}). 
{\sl 
\begin{df} %[$T$-weak Weyl relation] 
\label{df:T-weakWR} {\em 
: Let ${\cal H}$ be a Hilbert space, 
$T$ be a symmetric operator on ${\cal H}$, 
and $H$ be a self-adjoint operator on ${\cal H}$. 
If, for any $ \psi \in \mbox{\rm Dom}(T)$ 
and for any $ t \in {\bf R}^1 $, the relations 
$e^{-itH} \psi \in \mbox{\rm Dom}(T)$ and 
\begin{equation}
 Te^{-itH}\psi = e^{-itH} (T+t) \psi 
 \label{eqn:1.4}
\end{equation}
hold, then 
a pair of operators $T$ and $H$ 
is said to satisfy 
the $T$-weak Weyl relation ($T$-weak WR), 
or $T$ ($H$) is said to satisfy the $T$-weak WR with $H$ ($T$). 

} 
\end{df}
}
\noindent One can find, from the above definition, that 
$\mbox{\rm Dom}(Te^{-itH}) = \mbox{\rm Dom}(T)$, 
$\forall t \in {\bf R}^1 $. 
Thus the $T$-weak WR are represented merely by 
\begin{equation}
 Te^{-itH} = e^{-itH} (T+t),~~~\forall t \in {\bf R}^1 . 
 \label{eqn:1.5}
\end{equation}
It will follow that the time operator $T_0$, in 
Eq.\ (\ref{eqn:1.1}), 
is a symmetric operator on $L^2 ({\bf R}^1)$ 
and one of its symmetric extensions,  
denoted by $\widetilde{T_0}$, 
satisfies the $\widetilde{T_0}$-weak WR with $H_0$ 
(see Sec.\ \ref{sec:3}). 
%Thus in order to understand 
Thus as long as to see 
the qualitative properties of $T_0$ (or $\widetilde{T_0}$), 
it may suffice to examine the $T$-weak WR and the operators 
$T$ and $H$ satisfying this relation, 
by paying a particular attention to  
their spectra and to the uncertainty relation between them. 
%we consider to characterize the operator $T_0$, 
%by making use of the algebraic relation, $T_0$-weak WR. 
%we would see, 
%from the use of the algebraic relation, $T_0$-weak WR,  
%that the operator $T_0$ can be  
%characterized by the properties, which are  
%independent of the explicit form of $T_0$. 
We have obtained the fact that the time operator 
is deeply connected to the survival probability.   
Indeed, if a pair of operators $T$ and $H$ 
satisfies the $T$-weak WR,  the following inequality 
\begin{equation}
 \frac{4 \left( \Delta T\right)_{\psi} ^2  \| \psi \| ^2 }{t^2}
 \geq 
 \left| \left< \psi , e^{-itH } \psi   \right> \right| ^2 
 \label{eqn:survival1}
\end{equation} 
holds for every 
$\psi \in \mbox{\rm Dom}(T) $ and 
for every $t \in {\bf R}^1 \backslash \{ 0 \}$, 
where  
$\left( \Delta T\right)_{\psi}$ is the standard deviation 
of $T$ with respect to $\psi$, and 
$\left| \left< \psi , e^{-itH } \psi   \right> \right| ^2 $ 
is the survival probability of $\psi$ at time $t$. 
This is shown in Theorem\ \ref{th:survival}. 
As an application of this inequality, we have 
Corollary\ \ref{co:Hpp0} which states 
that $H$ has no point spectrum. 
Furthermore, it is shown that $H$ is absolutely continuous 
\cite{ka}, as in Theorem\ \ref{th:Hpp0}. 
This means 
that the existence of the time operator, which 
satisfies the $T$-weak WR with the 
Hamiltonian for some system, 
infers that 
%the point spectrum of $H$ is necessarily empty, and 
the system consists of only scattering states. 
Also, in Theorem\ \ref{th:uncertainty}, 
the absence of 
minimum-uncertainty states, for the uncertainty relation 
between $T$ and $H$,  
is proved, under some condition 
satisfied by the operators $\widetilde{T_0}$ and $H_0$.

In Sec.\ \ref{sec:2}, the connection 
among the CCR, Weyl relation, and $T$-weak WR 
is mentioned. 
Section \ref{sec:3} 
is devoted to the brief study of the Aharonov-Bohm 
time operator in Eq.\ (\ref{eqn:1.1}), 
to see a sign of the deep connection 
between the operator $T$ and the survival probability, 
followed by several statements in Sec.\ \ref{sec:4}. 
They include the inequality\ (\ref{eqn:survival1}) 
and the 
spectral properties of both $T$ and $H$, e.g.  
Theorem\ \ref{th:Hpp0}.  
Theorem\ \ref{th:uncertainty} is proved 
in Sec.\ \ref{sec:5}. 
Further discussion about the time operator 
is developed in Sec.\ \ref{sec:6}, 
on the basis of the results of the preceding sections 
and of the theory of  Schr\"{o}dinger operators. 
We mention other quantum systems than the 1DFPS 
for which an operator $T$ exists, 
to satisfy the $T$-weak WR with the Hamiltonian. 
In fact, for a certain class of quantum systems, 
time operators are easily constructed 
by unitary transformations of $\widetilde{T_0}$. 
%Existence of so many quantum systems for which a time operator 
%is allowed 
%is a surprising result, 
%although the construction of the time operator for these systems 
%is so easy. 
%
Concluding remarks are given in Sec.\ \ref{sec:7}.

%\newpage

\section{The canonical commutation relation, 
Weyl relation, and $T$-weak Weyl relation} 
\label{sec:2}

\setcounter{df}{0}

The $T$-weak WR in Eq.\ (\ref{eqn:1.4}) or (\ref{eqn:1.5}) 
is characterized more clearly, 
in the Heisenberg picture. 
The $T$-weak WR is represented, 
in an alternative form, as 
\begin{equation}
 T_t = T + tI , ~~~\forall t \in {\bf R}^1 ,
 \label{eqn:3.1}
\end{equation}
where $T_t :=e^{itH}Te^{-itH}$. 
It is now clear that $T$, which satisfies the $T$-weak WR with $H$, 
is shifted proportionally to the time parameter $t$ 
in the Heisenberg picture. 
This fact bring us an image of time for $T$. 
We also see, from this form, 
that $T$ is necessarily unbounded. 
It is, however, noted that in our investigation 
the $T$-weak WR in Eq.\ (\ref{eqn:1.5}) is more convenient 
than in Eq.\ (\ref{eqn:3.1}). 
The connection among the Weyl relation (WR) \cite{re}, 
the CCR and the $T$-weak WR  is very important, 
when one considers whether a symmetric operator $T$, satisfying 
the $T$-weak WR with the Hamiltonian for some system, is 
the time operator. Recall that the latter is defined 
as a symmetric operator 
satisfying the CCR with the same Hamiltonian as 
in Eq.\ (\ref{eqn:1.2}). 
In this respect, we put forward the next proposition.

{\sl 
\begin{pp} \label{pp:T-weakWRtoCCR}{\sl : 
Let ${\cal H}$ be a Hilbert space, $T$ be a closed symmetric 
operator on ${\cal H}$, 
and $H$ be a self-adjoint operator on ${\cal H}$.
If a pair of operators $T$ and $H$ satisfies the $T$-weak WR, 
then there is a dense subspace 
${\cal D} \subset {\cal H}$ such that 
\begin{flushleft}
\begin{tabular}{ll}
       (i) & ${\cal D} \subset \mbox{\rm Dom}(TH)\cap \mbox{\rm Dom}(HT),$ \\
       (ii) & $H:{\cal D} \rightarrow {\cal D},$ \\
       (iii) & The CCR holds in the meaning of that 
       $TH - HT = i $ on 
        $ \mbox{\rm Dom}(TH) 
       \cap \mbox{\rm Dom}(HT).$  
\end{tabular}
\end{flushleft} 
Moreover, if $T$ is self-adjoint, then the operators 
$T$ and $H$ satisfy the WR,  
\begin{equation}
e^{-isT} e^{-itH} = e^{-ist} e^{-itH}e^{-isT},~~~
\forall s,~ \forall t \in {\bf R}^1 .
\label{eqn:WR}
\end{equation}
}
\end{pp}}
\noindent The above (i), (ii), and (iii) are 
proved in the same manner as in the proof\ \cite{re-2}, 
by noting the strong continuity of $Te^{-itH}\psi$, 
$\forall \psi \in \mbox{\rm Dom}(T)$, 
by virtue of the $T$-weak WR, and 
the closedness of $T$, 
and also by considering the subspace spaned by  
the following subset of ${\cal H}$, 
as a subspace ${\cal D}$ in this proposition,  
\[
\left\{ \psi_f \in {\cal H} ~\left| ~ \psi_f := 
\int_{-\infty} ^{~\infty} f(s) e^{-isH} \psi ds,~ \forall 
f \in C_0 ^{\infty } ({\bf R}^1 ) \mbox{ and }
\forall \psi \in \mbox{\rm Dom}(T) \right. \right\} ,
\] 
where the integral is defined by Riemann's sense 
and thus a strong limit. 
The last part of the proposition is proved as follows.  
In the case of $T$ being self-adjoint, we see, 
from the $T$-weak WR (\ref{eqn:1.5}), 
that $\forall \phi \in {\cal H}$ and 
$\forall \psi \in \mbox{\rm Dom}(T)$, 
\begin{eqnarray}
\int_{{\bf R}^1} \lambda 
     d \left< \phi , e^{itH} F(\lambda) e^{-itH} \psi \right> 
& = & \left< \phi , e^{itH} T e^{-itH} \psi \right> 
 = \left< \phi , (T +t) \psi \right> \nonumber \\ 
& = & \int_{{\bf R}^1} (\lambda +t ) 
     d \left< \phi , F(\lambda)  \psi \right> 
 =  \int_{{\bf R}^1} \lambda 
     d \left< \phi ,  F_t (\lambda)  \psi \right> ,
      \nonumber 
\end{eqnarray}
where $\{ F(B)~|~B \in {\bf B}^1 \}$ 
is the spectral measure of $T$, 
${\bf B}^1$ is the $\sigma$-field which is generated by all open 
sets of ${\bf R}^1$, 
and $F_t (B) : =  F( \{ \lambda -t~|~\lambda \in B \} )$. 
%and $[B-t] := \{ \lambda -t~|~\lambda \in B \}$. 
From the uniqueness of the spectral resolution, 
%of self-adjoint 
%operators, 
this means  that 
$e^{itH} F(B) e^{-itH} = F_t (B)$,  
for all $t \in {\bf R}^1$. 
%
%
%that 
%$e^{itH} F(B) e^{-itH} = F([B-t])$,  
%$\forall t \in {\bf R}^1$, 
%by making use of 
%the $T$-weak WR and 
%the uniqueness of the spectral resolution of self-adjoint 
%operators, 
%where $\{ F(B)~|~B \in {\bf B}^1 \}$ 
%is the spectral measure of $T$, 
%${\bf B}^1$ is the $\sigma$-field which is generated by all open 
%sets of ${\bf R}^1$, 
%and $[B-t] := \{ \lambda -t~|~\lambda \in B \}$. 
Then it follows that $\forall \psi \in {\cal H}$ and 
$\forall s \in {\bf R}^1$, 
\begin{eqnarray}
\left< \psi , e^{itH} e^{-isT} e^{-itH} \psi \right> 
& = & \int_{{\bf R}^1} e^{-is\lambda } 
     d \left< \psi , e^{itH} F(\lambda) e^{-itH} \psi \right> 
 =  \int_{{\bf R}^1} e^{-is\lambda } 
     d \left< \psi ,  F_t (\lambda)  \psi \right> \nonumber \\ 
& = & \int_{{\bf R}^1} e^{-is( \lambda + t) } 
     d \left< \psi ,  F(\lambda)  \psi \right> 
= \left< \psi , e^{-ist } e^{-isT }  \psi \right> . \nonumber
\end{eqnarray}
By using the polarization identity, 
we can obtain the WR (\ref{eqn:WR}). 
According to von Neumann's uniqueness theorem, 
with respect to the solution of the WR \cite{re}, 
we had better to define $T$, 
which appears in the $T$-weak WR, 
as a symmetric operator, 
to allow the operator $H$ (corresponding to the Hamiltonian) 
to be bounded from below. 
We note here that if a symmetric operator $T$ satisfies 
the $T$-weak WR 
with some self-adjoint operator $H$, 
then the closure of $T$, denoted by $\overline{T}$, 
also  satisfies the $\overline{T}$-weak WR with the same $H$. 
This is easily verified by usual calculation. 
It is guaranteed, from this proposition, that 
a symmetric operator $T$, satisfying the $T$-weak WR 
with the Hamiltonian for some system, is the time operator, 
and thus it is significant to examine the $T$-weak WR 
in the general analysis of the time operator. 
As a summary,  
we remark again that the following relations, 
\[ \mbox{ WR } \Rightarrow  \mbox{ $T$-weak WR } 
\Rightarrow  \mbox{ CCR } \]
%\begin{center}
%\begin{tabular}{lll}
%   & $e^{-isT} e^{-itH} = e^{-ist} e^{-itH}e^{-isT}$ \hspace{20mm} 
%   & (WR) \\
%   $\Rightarrow $ & $Te^{-itH} = e^{-itH} (T+t)$ 
%   & ($T$-weak WR) \\
%   $\Rightarrow $ & $TH-HT \subset i $ & (CCR)~~,
%\end{tabular}
%\end{center}
%
hold, in the sense of Proposition\ \ref{pp:T-weakWRtoCCR}, 
even though, in general the converses do not hold, 
as is already mentioned in Sec.\ \ref{sec:1}.

%\newpage

\section{The Aharonov-Bohm time operator } 
\label{sec:3} 

\setcounter{df}{0}

%
%
%\section{The important role of the $T$-weak WR 
%in the analysis of the time operator } 
%\label{sec:3} 

\setcounter{df}{0}

Let us consider the Hilbert space $L^2 ({\bf R}^1)$. 
The operator $T_0$ on $L^2 ({\bf R}^1)$ in Eq.\ (\ref{eqn:1.1}), 
\begin{eqnarray*}
 T_0 & := & \frac{1}{2}\left( QP^{-1}~+~P^{-1} Q \right),  
\end{eqnarray*}
is defined in its domain 
%\begin{eqnarray*}
$ \mbox{\rm Dom}(T_0) := %& := & 
 \mbox{\rm Dom}(QP^{-1}) \cap \mbox{\rm Dom}(P^{-1} Q )$, %\\ 
%\end{eqnarray*}
where $P$ is the momentum operator on $L^2 ({\bf R}^1)$ 
for the 1DFPS, 
and $P^{-1}$ its inverse. 
In the axiomatic quantum mechanics, $P$ is defined as 
$P:=-iD_x$, where $D_x $ is a differential operator on 
$L^2 ({\bf R}^1)$, and its domain consists of 
the $L^2$-functions which belong to $AC({\bf R}^1)$, and 
%and square integrable 
%functions on ${\bf R}^1$ such that 
satisfy that their derivatives are also 
included in $L^2 ({\bf R}^1)$ \cite{ak}. 
$AC(\Omega)$ ($\Omega$ is an open set of ${\bf R}^1$) 
is the set of functions on $\Omega$, 
which are absolutely continuous on 
all bounded closed intervals of $\Omega$. 
The free Hamiltonian $H_0$ 
for this system is $H_0 := P^2 /2 $. 
The position operator $Q$ on $L^2 ({\bf R}^1)$ 
is defined as an operator of multiplication 
by $x$ on $L^2 ({\bf R}^1 )$, 
denoted by $M_{x}$, and its domain 
consists of $L^2$-
%square integrable 
functions, defined by $\psi$, such that $
\displaystyle{
 \int_{{\bf R}^1} \left| x \psi (x) \right| ^2 dx 
} $ is finite. 
It is noted that in the definition of $T_0$, 
$P^{-1}$ is well defined and becomes a self-adjoint operator 
on $L^2 ({\bf R}^1)$. 
This is because, for any self-adjoint operator $A$,  
if its inverse $A^{-1}$ exists, 
$A^{-1}$ should be self-adjoint \cite{ak-2}.  
In our case, $P^{-1}$ exists 
since $P$ is an injection, i.e., %from the fact 
$\mbox{\rm Ker}(P^{-1}) =\{ 0\}$, where 
$\mbox{\rm Ker}(A) := \{\psi \in \mbox{\rm Dom}(A)~|~ A\psi =0 \}$. 

In the momentum representation of $T_0$, 
we have 
\[ FT_0 F^{-1} = \frac{1}{2}\left( iD_k M_{1/k} + M_{1/k} iD_k \right) \] 
and its domain 
\begin{eqnarray}
 \mbox{\rm Dom}(FT_0 F^{-1})& = & \mbox{\rm Dom}(D_k M_{1/k}) 
          \cap \mbox{\rm Dom}(M_{1/k} D_k)  \nonumber \\
      &=& \left\{ \psi \in \mbox{\rm Dom}(M_{1/k} )  
      ~\left|~  M_{1/k} \psi \in 
      \mbox{\rm Dom}(D_k ) \right. \right\}  \nonumber \\ 
      & & 
      \cap \left\{ \psi \in \mbox{\rm Dom}(D_k ) 
      ~\left|~  D_k \psi \in 
       \mbox{\rm Dom}(M_{1/k} ) \right. \right\} ,
       \label{eqn:Dom(FT_0F^{-1})}
\end{eqnarray}
where $F$ is the Fourier transformation from 
$L^2 ({\bf R}^1)$ onto $L^2 ({\bf R}_k ^1)$, 
%Winter, JMP, 1998. 
and the use has been made of 
the relations $FQ F^{-1} = iD_k$, $FP F^{-1} = M_k$, 
and $FP^{-1} F^{-1} = M_{1/k}$. 
At first sight, $\mbox{\rm Dom}(T_0 )$ 
seems to be rather restricted, 
because of the existence of $P^{-1}$ in the definition of $T_0$. 
The following simple example by\ Kobe \cite{ko} 
may be considered to support this anticipation. 
%We can recognize this fact actually from the next example \cite{ko}, 
%described in the momentum representation. 
%
%
{\sl 
\begin{ex} \label{ex:1} {\em : 
Let us consider the functions  
$\phi _n (k) := k^n N_n e^{-a_0 k^2 } \in  L^2 ({\bf R}_k ^1)$, 
where 
$n \in {\bf Z}$, $n \geq 0$, $a_0 >0 $ 
and $N_n$ is a normalization factor. 
We see that for any integer $n \geq 2$, 
$\phi _n  \in \mbox{\rm Dom}(FT_0 F^{-1}) $. 
The action of $FT_0 F^{-1}$ on each $\phi _n ~( n \geq 2 )$ 
is, by direct calculation, % and for these functions, 
%\begin{equation}
\[
       FT_0 F^{-1} \phi _n (k)  =  
\frac{i}{2}\left[ (2n-1)k^{n-2} -2a_0 k^{n-1} 
-2a_0 k^n \right] N_n e^{-a_0 k^2 } ~~. 
\]
%\label{eqn:2.0}
%\end{equation}
In the case of $n=0,~1$, however,  
the right-hand side of the above equation 
is formally not square integrable, and  
thus $\phi _0 ,~ \phi_1 \notin \mbox{\rm Dom}(FT_0 F^{-1})$. 
 }
\end{ex}
}
Notice that in spite of this example, $\mbox{\rm Dom}(T_0)$ 
is dense in $L^2 ({\bf R}^1)$. 
This can be seen from the fact that the subspace 
${\cal C}_{\rm i}$ is included in  
$\mbox{\rm Dom}(FT_0 F^{-1}) $ and 
is dense in $L^2 ({\bf R}_k ^1)$. 
${\cal C}_{\rm i} $ is defined as 
\begin{equation}
 {\cal C}_{\rm i} := \left\{ \psi \in C^{\infty }_0 ({\bf R }_k ^1 )~
\left| ~ \mbox{supp} ~\psi  \subset {\bf R}_k ^1 \backslash \{ 0\} \right. 
\right\} ,
\label{eqn:Ci} 
\end{equation}
where 
%$C^{\infty }_0 ({\bf R }_k ^1 )$ is a set of infinitely 
%differentiable functions, whose supports are bounded, and 
$\mbox{supp}\ \psi $ denotes the support of $\psi $,  
i.e. the closure of $\{ k \in {\bf R }_k ^1 ~|~ \psi (k) \neq 0 \}$. 
Therefore the adjoint operator of $T_0$, denoted by $T_0 ^*$, 
can be defined. 
Then, $T_0$ is symmetric, because 
%of the use of the following relation that 
%\begin{equation}
\[ T_0 ^* \supset  \frac{1}{2}\left( (QP^{-1})^* + (P^{-1} Q)^* \right) 
       \supset  \frac{1}{2}\left( (P^{-1})^* Q^* + Q^* (P^{-1})^* \right) 
       = T_0,   
%\end{equation} 
\] 
where we have used the fact that 
$Q^* =Q$ and $(P^{-1})^* =P^{-1}$. 
It is noted that $T_0$ and $H_0$ 
do not satisfy the $T_0$-weak WR,  
$T_0 e^{-itH_0}  = e^{-itH_0} (T_0 + t)  ,
~~ \forall t \in {\bf R}^1 $ . 
%( $\mbox{{\rm Im}~} t \leq 0$ )   . 
%\begin{equation}
%\[ 
% T_0 e^{-itH_0} \psi = e^{-itH_0} (T_0 + t) \psi ,  
% \label{eqn:T_0weakWR}
%\end{equation} 
%~~~\forall \psi \in \mbox{\rm Dom}(T_0),~~ \forall t \in {\bf C} 
%( \mbox{{\rm Im}~} t \leq 0).   
%\]
Because $\mbox{\rm Dom}(FT_0F^{-1})$ in Eq.\ 
(\ref{eqn:Dom(FT_0F^{-1})}) is not 
invariant under the action of $e^{-itM_{k^2 /2 }}$ 
for all $t \neq 0$, that is  for any $t \neq 0$, 
there is some vector $\psi \in \mbox{\rm Dom}(FT_0F^{-1})$ 
satisfying 
$e^{-itM_{k^2 /2 }} \psi \notin \mbox{\rm Dom}(FT_0F^{-1})$. 
For instance, consider the following $L^2$-function $g$ of 
$k \in {\bf R}_k ^1 $, 
\[g(k) :=\left\{ \begin{array}{cc}
                e^{-1/k^2 } \displaystyle{ \frac{1}{1+|k|^s } } 
                &~~(k\neq 0) \\ 
                0       & ~~(k = 0)
                 \end{array}
                \right.  \]
%It is ssen that 
where $1/2 < s \leq 3/2 $. 
Then $g$ is $C^{\infty}$-function. 
One can see that $g \in \mbox{\rm Dom}(FT_0F^{-1})$, however, 
$e^{-itM_{k^2 /2}}g \notin \mbox{\rm Dom}(FT_0F^{-1})$, 
$\forall t \neq 0$. 
This follows from the fact that 
$e^{-itM_{k^2 /2}}g \notin \mbox{\rm Dom}(D_k)$, 
$\forall t \neq 0$. 
We here introduce a symmetric extension of 
$T_0$ on $L^2 ({\bf R}^1)$, 
denoted by $\widetilde{T_0}$, which will   
satisfy the $\widetilde{T_0}$-weak WR with $H_0$, 
and is defined, in the momentum representation, as follows, 
%This operator is defined, in the momentum representation, as   
\[
\mbox{\rm Dom}(F\widetilde{T_0}F^{-1}) := \left\{ 
              \psi \in L^2 ({\bf R}_k ^1) ~\left| ~
              \begin{array}{c}
              \psi \in AC({\bf R}_k ^1 \backslash \{ 0 \} ), ~~
              \displaystyle{  \lim_{k \rightarrow 0} 
              \frac{\psi (k)}{|k|^{1/2}} =0 }, \mbox{ and}\\
              \displaystyle{  
              \int_{{\bf R}_k ^1 \backslash \{ 0 \} }
              \left| \frac{d\psi (k) /k}{dk} + \frac{1}{k}
              \frac{d\psi (k)}{dk} \right|^2  dk < \infty 
              }
              \end{array} 
              \right. 
              \right\} , 
\] 
and its action, 
%on $\psi \in \mbox{\rm Dom}(F\widetilde{T_0}F^{-1})$ is 
\begin{equation}
  F\widetilde{T_0} F^{-1} \psi (k) = \frac{i}{2} \left( 
  \frac{d\psi (k) /k}{dk} + \frac{1}{k}
  \frac{d\psi (k)}{dk}  \right) ,~~~
  \mbox{ a.e. } k \in {\bf R}_k ^1 \backslash \{ 0 \} ,~~
  \forall \psi \in \mbox{\rm Dom}(F\widetilde{T_0}F^{-1}) . 
  \label{eqn:tilde{T_0}}
\end{equation}
It is seen that $\mbox{\rm Dom}(F\widetilde{T_0}F^{-1}) $ 
is a subspace of $L^2 ({\bf R}_k ^1 )$, and 
$F\widetilde{T_0}F^{-1}$ is a linear operator on 
$L^2 ({\bf R}_k ^1 )$. 

{\sl
\begin{pp} \label{pp:rapidly} {\sl 
: $\widetilde{T_0} $ is a symmetric extension of $T_0$. 
}
\end{pp}
}

{\sl Proof} : 
$F\widetilde{T_0} F^{-1} $ being symmetric follows from 
that $\forall \psi ,~\forall \phi \in 
\mbox{\rm Dom}(F\widetilde{T_0}F^{-1}) $,  
\begin{eqnarray*}
& & \int_{(0, \infty ) } 
\bar{\phi} (k)  \frac{i}{2}\left( 
\frac{d\psi (k) /k}{dk} + \frac{1}{k}
\frac{d\psi (k)}{dk} \right)  dk - 
\int_{(0, \infty ) } 
\frac{-i}{2}\left( 
\frac{d\bar{\phi} (k) /k}{dk} + \frac{1}{k}
\frac{d\bar{\phi} (k)}{dk} \right) \psi (k)   dk  \\
& = & i \left(  
\lim_{b \rightarrow \infty} \frac{\bar{\phi} (b) \psi (b)}{b} - 
\lim_{a \downarrow 0} \frac{\bar{\phi} (a) \psi (a)}{a} 
\right)  =  0 ,   
\end{eqnarray*}
%The last equality is brought from  
where 
$\lim_{b \rightarrow \infty} \bar{\phi} (b) \psi (b) / b =0 $ and 
$\lim_{a \downarrow 0} \bar{\phi} (a) \psi (a) /a =0 $ 
are used. The former is brought from  the integrability of 
$\bar{\phi} (k) \psi (k)$, and the latter from the boundary 
conditions of $\bar{\phi} (k)$ and $\psi (k)$ at the origin. 
By considering the left half-line in the same manner, 
we can obtain that  $\forall \psi ,~\forall \phi \in 
\mbox{\rm Dom}(F\widetilde{T_0}F^{-1}) $, 
$\left< \phi, F\widetilde{T_0}F^{-1} \psi  \right> 
= \left< F\widetilde{T_0}F^{-1} \phi,  \psi  \right> $, 
that is, $F\widetilde{T_0}F^{-1}$ is symmetric. 
To see that $\widetilde{T_0}$ is an extension of $T_0$, 
i.e. $\widetilde{T_0} \supset T_0$,  
it is sufficient that every $\psi \in \mbox{\rm Dom}(FT_0 F^{-1}) $ 
satisfies the boundary condition at the origin, which appears 
in the definition of $\mbox{\rm Dom}(F\widetilde{T_0}F^{-1}) $.  
This is easily verified as follows.  
Consider a $\psi \in \mbox{\rm Dom}(FT_0 F^{-1}) $ 
in Eq.\ (\ref{eqn:Dom(FT_0F^{-1})}) , 
then $\psi (k) /k$ 
belongs to $AC({\bf R}_k ^1)$, and thus, 
$\lim_{k \rightarrow 0} \psi (k) /k$ exists.  
Thus $\lim_{k \rightarrow 0} |\psi (k) /|k|^{1/2} | 
= \lim_{k \rightarrow 0} |k|^{1/2} |\psi (k) /k | =0$. 
\hfill \qed

\noindent
This operator may be more understood, from the view of the 
energy representation which 
was emphasized by\ Egusquiza and\ Muga, 
and many other authors  
(see \cite{eg} and the references therein).   

It is, now, noted that  
$\mbox{\rm Dom}(\widetilde{T_0})$ is an invariant subspace 
of $e^{-itH_0}$. Because, for every 
$\psi \in \mbox{\rm Dom}(F\widetilde{T_0} F^{-1})$ and 
$t \in {\bf C} $  ( $\mbox{{\rm Im}~} t \leq 0 $), 
$\lim_{k \rightarrow 0} e^{-itk^2 /2 } \psi (k) /|k|^{1/2} =0$, 
and for almost everywhere $k \in {\bf R}_k ^1 \backslash \{ 0 \}$, 
\[ 
\frac{i}{2} \left( 
\frac{d e^{-itk^2 /2 } \psi (k) /k}{dk} + \frac{1}{k}
\frac{d e^{-itk^2 /2 } \psi (k)}{dk}  \right) 
= t e^{-itk^2 /2 } \psi (k) + 
e^{-itk^2 /2 } \frac{i}{2} \left( 
\frac{d  \psi (k) /k}{dk} + \frac{1}{k}
\frac{d  \psi (k)}{dk}  \right) 
,~~~
%\mbox{ a.e. } k \in {\bf R}_k ^1 \backslash \{ 0 \} , 
\]
where the right-hand side is square-integrable. 
Therefore $e^{-it M_{k^2 /2}} \psi $ 
is included in $\mbox{\rm Dom}(F\widetilde{T_0 }F^{-1})$, 
and, as a result,  
$\widetilde{T_0}$ can satisfy the $\widetilde{T_0}$-weak WR 
with $H_0$, 
\begin{equation}
 \widetilde{T_0} e^{-itH_0}  = e^{-itH_0} (\widetilde{T_0} + t)  , ~~~ 
% \forall \psi \in \mbox{\rm Dom}(\widetilde{T_0}) ,~~
 \forall t \in {\bf C} ~ 
( \mbox{{\rm Im}~} t \leq 0 ),     
 \label{eqn:T_0weakWR}
\end{equation} 
whereas $T_0$ does not satisfy the $T_0$-weak WR with $H_0$. 
%
%This relation is easily shown in the momentum representation, 
%as in Eq.\ (\ref{eqn:2.0}), 
%by the virtue of the form of $\mbox{\rm Dom}(\widetilde{T_0})$ 
%and through the inner product between 
%$\phi \in {\cal C}_i $ and $\widetilde{T_0} e^{-itH_0} \psi, ~
%\psi \in \mbox{\rm Dom}(\widetilde{T_0})$.  
It is seen that $\widetilde{T_0}$ is not self-adjoint. 
Because, if it was so, $\widetilde{T_0}$ and $H_0$ would have to satisfy 
the WR from Proposition\ \ref{pp:T-weakWRtoCCR}.  
The latter is however in contradiction to 
the nonnegativity $H_0 \geq 0$.

Consider the subspace ${\cal C}_{\rm i} $ in Eq.\ (\ref{eqn:Ci}). 
It is easily 
seen that ${\cal C}_{\rm i}$ is an invariant subspace 
of $F\widetilde{T_0} F^{-1}$, that is, 
$F\widetilde{T_0} F^{-1} :~{\cal C}_{\rm i} \rightarrow 
{\cal C}_{\rm i} $. 
Thus $F\widetilde{T_0} F^{-1}$ can act any times on 
${\cal C}_{\rm i} $. 
In the position representation, 
this property is described as 
$\widetilde{T_0} :~F^{-1} {\cal C}_{\rm i} \rightarrow F^{-1} {\cal C}_{\rm i} $ 
where $ F^{-1} {\cal C}_{\rm i} := 
\{ \psi \in L^2 ({\bf R}^1) ~|~\psi = F^{-1} \eta ,~ 
\eta \in {\cal C}_{\rm i} \}$. 
${\cal C}_{\rm i} $ may be regarded as an important subspace which 
determines the property of $\widetilde{T_0}$.
Indeed, using the $\widetilde{T_0}$-weak WR in Eq.\ (\ref{eqn:T_0weakWR}), 
we can obtain the following statement: 
{\sl
\begin{pp} \label{pp:rapidly} {\sl 
: For any nonnegative integer $n,~m$ and for any $\psi , \phi  
\in {\cal C}_{\rm i} $, 
\[ \lim_{t \to \pm \infty } |t|^n 
\left| \frac{d^m \left< \phi , e^{-it H_0 } \psi   \right> }{d t^m}
\right| =0 ~. \]
That is, 
the probability amplitude, 
$\left< \phi , e^{-it H_0 } \psi   \right> $, is 
a rapidly decreasing function of $t \in {\bf R}^1 $.}
\end{pp}
}

{\sl Proof} : Let $\psi , \phi \in F^{-1} {\cal C}_{\rm i}$. 
Since $F^{-1} {\cal C}_{\rm i}$ is an invariant subspace of $\widetilde{T_0}$, 
%we see that 
thus $\widetilde{T_0} \phi, \widetilde{T_0} \psi \in F^{-1} {\cal C}_{\rm i} $. 
%\subset \mbox{\rm Dom}(\widetilde{T_0})$. 
By using Eq.\ (\ref{eqn:T_0weakWR}), we have 
%\begin{eqnarray}
%      \left< \phi , e^{-it H_0 } \widetilde{T_0} \psi   \right> & = & \left< 
%      \phi , (\widetilde{T_0} -t ) e^{-it H_0 } \psi   \right>  
%       \nonumber \\ 
%       & = & \left< \widetilde{T_0} \phi , e^{-it H_0 } \psi   \right> 
%       -t \left< \phi , e^{-itH_0 } \psi    \right>. 
%       \label{eqn:2.2}
%\end{eqnarray}
\begin{equation}
     \left< \phi , e^{-it H_0 } \widetilde{T_0} \psi   \right>  =  \left< 
      \phi , (\widetilde{T_0} -t ) e^{-it H_0 } \psi   \right>  
        =  \left< \widetilde{T_0} \phi , e^{-it H_0 } \psi   \right> 
       -t \left< \phi , e^{-itH_0 } \psi    \right>,  
      \label{eqn:2.2}
\end{equation}
where $\left< \cdot , \cdot \right>$ denotes the inner product 
in $L^2 ({\bf R}^1)$. 
Note that 
$\forall \psi \in L^2 ({\bf R}^1)$, 
$\mbox{w-}\lim_{t \to \pm \infty } e^{-it H_0 } \psi =0$ \cite{re3-3}, 
because 
$H_0$ is (spectrally) absolutely continuous \cite{ka}. 
This means that $ \lim_{t \to \pm \infty }
 \left< \phi , e^{-it H_0 } \widetilde{T_0} \psi   \right>
  = \lim_{t \to \pm \infty } \left< \widetilde{T_0} \phi , 
  e^{-it H_0 } \psi   \right> =0  $, 
which leads to the relation 
 $ \lim_{t \to \pm \infty } t \left< \phi , e^{-it H_0 } \psi  
  \right> =0 $. 
In order to show that 
for any integer $n \geq 2$ and 
for any $\psi , ~\phi \in {\cal C}_{\rm i} $, 
$ \lim_{t \to \pm \infty } t^n 
\left< \phi , e^{-it H_0 } \psi \right> =0 $,  
we observe that  
$ \widetilde{T_0} ^{~k} : F^{-1} {\cal C}_{\rm i} \rightarrow 
F^{-1} {\cal C}_{\rm i} $, 
and that the following relations similar to Eq.\ (\ref{eqn:2.2}) 
hold for every $n \in {\bf N}$:  
\begin{eqnarray*}
       \left< \phi , e^{-it H_0 } \widetilde{T_0} ^{~n+1} \psi   \right> & = & 
       \left< \phi , (\widetilde{T_0} -t)^{~n+1} e^{-it H_0 } \psi   \right>  \\
       & = & \sum _{k=0} ^{n} (-t)^k {{n+1}\choose{k} }
       \left<\widetilde{T_0}
 ^{~n+1-k} \phi ,  e^{-it H_0 } \psi   \right> 
 +(-t)^{n+1} \left<  \phi , e^{-it H_0 } \psi   \right> ~.
\end{eqnarray*}
Then $ \lim_{t \to \pm \infty } t^n 
\left< \phi , e^{-it H_0 } \psi   \right> =0 $ 
is proved recursively for any integer $n \geq 2$.
In order to show that  
$\left< \phi , e^{-it H_0 } \psi   \right>$ 
is infinitely differentiable on ${\bf R}^1 $, 
it is sufficient to use the fact that  
$\forall \psi  \in F^{-1} {\cal C}_{\rm i} $, 
$e^{-it H_0 } \psi $ 
is infinitely and strongly differentiable on ${\bf R}^1 $, 
and $F^{-1} {\cal C}_{\rm i} $ is also an invariant subspace of $H_0$, 
that is, 
$H_0 : F^{-1} {\cal C}_{\rm i} \rightarrow F^{-1} {\cal C}_{\rm i} $.  
\hfill \qed

We note that $\forall \psi , \phi  
\in {\cal C}_{\rm i} $, 
$\left< \phi , e^{-it H_0 } \psi \right> $ 
converges to $0$\ as $t \rightarrow \pm \infty $, rapidly than 
any inverse-power of $t$. This fact is not trivial and is seen 
from the next example.

{\sl
\begin{ex} {\em : 
 Define  the survival probability of $\psi$ 
 as a function of $t \in {\bf R}^1$, i.e. 
 $P_{\psi} (t) := 
 \left| \left< \psi , e^{-itH_0 } \psi   \right> \right| ^2 $, 
 where $\psi $ is an arbitrary element in $L^2 ({\bf R}_k ^1)$. 
 Then, for a particular 
 $\phi_n ,~n \geq 2 $ in Example\ \ref{ex:1}, 
 $ P_{\phi _n } (t) = \left( 1+ t^2 / 16 a_0 ^2  \right) ^{-n-1/2 } $ 
and this converges to $0$ as $t \rightarrow \pm \infty $ 
as, at most, a power function of $t$.
 }
\end{ex}
}

From the above statement and example, 
we may expect 
that there is a connection between 
$T_0$ (or $\widetilde{T_0}$) and the survival probability. 
This expectation is also inspired from the works done by 
Bhattacharyya \cite{bh}.

%\newpage

\section{Connection between the time operator 
and the survival probability} 
\label{sec:4}

\setcounter{df}{0}

%
%
%\section{Some implications of the $T$-weak WR} 
%\label{sec:4}

If we assume the existence of a symmetric operator $T$ 
which satisfies the $T$-weak WR with the Hamiltonian 
for some system, i.e. the time operator, 
several statements are derived, 
in a rigorous form, which concern to 
the connection between the time operator 
and the survival probability. 
Before deriving these statements, 
we introduce a few definitions. 
%
%In this section,  
%several statements about each of operators $T$ and $H$, 
%which satisfy the $T$-weak WR, are derived. 
%In particular, Theorems\ \ref{th:survival}, 
%\ref{th:Hpp0}, and \ref{th:uncertainty}, 
%shown here, turn out to be important %valuable 
%for characterizing the $T$-weak WR. 
%Before deriving these statements, 
%we introduce a few definitions. 
%{\sl
%\begin{df}$:$ {\em 
Let $T$ be a symmetric operator on the Hilbert space ${\cal H}$ 
and define
\begin{equation}
\left< T \right>_{\psi}  :=  \left< \psi , T \psi   \right> ,~~
\left( \Delta T\right)_{\psi}  :=  \left\| \left( T - \left< T \right>_{\psi} 
\right) \psi \right\| , ~~
\forall \psi \in \mbox{\rm Dom}(T), 
\label{eqn:std}
\end{equation}
where $\left< \cdot , \cdot \right>$ denotes the inner product 
in ${\cal H}$, and $\| \cdot \|$ the norm in ${\cal H}$, 
defined by this inner product.  
$\left< T \right>_{\psi} $ 
and $\left( \Delta T\right)_{\psi}$ 
are respectively called the expectation 
and standard deviation 
of $T$ with respect to the state $\psi$. 
%}
%\end{df}}

\begin{th} 
\label{th:survival} {\sl 
$:$ Let $T$ be a symmetric operator on ${\cal H}$, and 
$H$ be a self-adjoint operator on ${\cal H}$. 
Then if a pair of operators $T$ and $H$ 
satisfies the $T$-weak WR,  the inequality\ 
(\ref{eqn:survival1}) holds. 
} 
\end{th}

{\sl Proof}
: Let us define self-adjoint operators 
$\cos (tH) := \left( e^{itH} + e^{-itH } \right) /2$ 
and $\sin (tH ) := \left( e^{itH} - e^{-itH} \right) /2i$.  
Then, from the $T$-wea WR in Eq.\ (\ref{eqn:1.5}), 
we can obtain two commutation relations 
\begin{equation}
      \left[ T,~  \cos (tH) \right]   =   -it \sin ( tH) , ~~~ 
        \left[ T,~ \sin (tH)  \right]   =   it \cos(tH ) . 
        \label{eqn:survival2}
\end{equation}
From  
the above commutation relations, we can derive 
the uncertainty relations.  
From the first relation in Eq.\ (\ref{eqn:survival2}), we have 
that 
%\begin{equation}
\[
       \left( \Delta T \right) _{\psi} ^2  \| \cos (tH 
) \psi \|^2  \geq  \frac{t^2}{4}\left| \left< \psi ,
 \left[ T, \cos (tH) \right] \psi   \right> \right| ^2 
           =  \frac{ t^2 }{4}\left|  \mbox{{\rm Im}~} 
           \left< \psi , e^{-itH} \psi 
            \right> \right| ^2 ,~~
\]
%\end{equation}
%\begin{eqnarray*}
%       \left( \Delta T \right) _{\psi} ^2  \| \cos (tH 
%) \psi \|^2 & \geq & \frac{1}{4}\left| \left< \psi ,
% \left[ T, \cos (tH) \right] \psi   \right> \right| ^2 \\
%          & = & \frac{1}{4} \left| \left< \psi ,  -it \sin ( tH )
%\psi   \right> \right| ^2 \\
%          & = & \frac{ t^2 }{4}\left|  \mbox{{\rm Im}~} \left< \psi , e^{-itH} \psi 
%            \right> \right| ^2 ~.
%\end{eqnarray*}
$\forall \psi \in \mbox{\rm Dom}(T),~ \forall t \in {\bf R}^1 $. 
Similarly, the second relation in Eq.\ (\ref{eqn:survival2}) 
gives us an inequality  
\[ \left( \Delta T \right) _{\psi} ^2  \| \sin (tH ) \psi \|^2  \geq 
\frac{ t^2 }{4}\left|  \mbox{{\rm Re}~} \left< \psi , e^{-itH} \psi 
           \right> \right| ^2 ,~~
            \forall \psi \in \mbox{\rm Dom}(T),~
            \forall t \in {\bf R}^1 . 
\]
%
%\begin{eqnarray*}
%       \left( \Delta A \right) _{\psi} ^2  \| \sin (tH 
%) \psi \|^2 & \geq & \frac{1}{4}\left| \left< \psi , \left[ A , \sin
%(tB) \right] \psi   \right> \right| ^2 \\
%          & = & \frac{1}{4} \left| \left< \psi , it \cos ( tB )
%\psi   \right> \right| ^2 \\
%          & = & \frac{ t^2 }{4}\left| \Re \left< \psi , e^{-itB} \psi 
%            \right> \right| ^2  
%\end{eqnarray*}
%
%
Adding these inequalities together, and taking into account of 
the relation 
$ \| \cos (tH ) \psi \|^2 + \| \sin (tH ) \psi \|^2 =\| \psi \|^2 $, 
the inequality (\ref{eqn:survival1}) can be obtained. 
%we obtain 
%\[ \left( \Delta T \right) _{\psi} ^2 \| \psi \|^2 \geq \frac{t^2 }{4}
%\left| \left< \psi , e^{-itH } \psi   \right> \right| ^2 
%=\frac{t^2 }{4} P_{\psi } (t) .\]
~~ \hfill \qed

Two corollarys follow from 
the inequality\ (\ref{eqn:survival1}).

{\sl 
\begin{co} \label{co:noneigenvalue} {\sl : 
Let $T$ be a symmetric operator on ${\cal H}$, and 
$H$ be a self-adjoint operator on ${\cal H}$. 
Then if a pair of operators $T$ and $H$ 
satisfies the $T$-weak WR, 
$T$ has no point spectrum. }
\end{co}}

{\sl Proof} $:$ Suppose that there existed 
an eigenvector $\psi_0 \in \mbox{\rm Dom}(T)$ belonging to 
an eigenvalue $\lambda \in {\bf R}^1$ of $T$, 
that is, $T\psi_0 = \lambda \psi_0$ 
and $\| \psi_0 \| = 1$. 
Then we see that $\left( \Delta T\right)_{\psi_0} =0$, 
from the definition in Eq.\ (\ref{eqn:std}). 
It follows, from Theorem\ \ref{th:survival}, that 
$\left< \psi_0 , e^{-itH} \psi_0 \right> =0$, 
$\forall t \in {\bf R}^1 \backslash \{ 0\}$. 
Since $e^{-itH}$ is strongly continuous at any $t \in {\bf R}^1$, 
we have that $\| \psi_0 \|^2 =\lim_{t \rightarrow 0} 
\left< \psi_0 , e^{-itH} \psi_0 \right> 
= 0$, and this is in contradiction to the premise. 
Thus $T$ has no point spectrum. \hfill \qed

\begin{co} \label{co:Hpp0} {\sl 
$:$  Let $T$ be a symmetric operator on ${\cal H}$, and 
$H$ be a self-adjoint operator on ${\cal H}$. 
If a pair of operators $T$ and $H$ 
satisfies the $T$-weak WR, then  
$H$ has no point spectrum. 
%, that is,   
%${\cal H}_{\rm sing} (H) = {\cal H}_{\rm pp} (H) =\{ 0\}$. 
}
\end{co}

{\sl Proof} : %It is first seen that $H$ has no point spectrum, 
%i.e. ${\cal H}_{\rm pp} (H) =\{ 0\}$.  
Since $\mbox{\rm Dom}(T)$ is dense in ${\cal H}$, 
for each $\psi \in {\cal H}$ 
there is a sequence $\{ \psi_n \}_{n=1} ^{\infty } 
\subset \mbox{\rm Dom}(T)$, 
satisfying $\psi_n \rightarrow \psi,~~n \rightarrow \infty$. 
It follows that 
\begin{eqnarray*}
& & \left| \left< \psi , e^{-itH} \psi \right> 
-  \left< \psi_n , e^{-itH} \psi_n \right>  \right| \\ 
&=&  \left| \left< \psi , e^{-itH} \psi \right> 
- \left< \psi , e^{-itH} \psi_n \right> 
+ \left< \psi , e^{-itH} \psi_n \right>               
- \left< \psi_n , e^{-itH} \psi_n \right>   \right| \\
&\leq & \| e^{itH} \psi \| \| \psi - \psi_n \| 
+ \| \psi - \psi_n \| \| e^{-itH} \psi_n \| \\
&\leq & ( \| \psi \| + \| \psi_n \| ) \| \psi - \psi_n \|,  
\end{eqnarray*}
and thus, 
\begin{eqnarray*}
\limsup_{t \rightarrow \pm \infty } 
\left| \left< \psi , e^{-itH} \psi \right> \right| 
& \leq & 
\limsup_{t \rightarrow \pm \infty } 
\left| \left< \psi_n , e^{-itH} \psi_n \right> \right| 
+ ( \| \psi \| + \| \psi_n \| ) \| \psi - \psi_n \| \\
& = & ( \| \psi \| + \| \psi_n \| ) \| \psi - \psi_n \| ,  
\end{eqnarray*}
where we use the inequality (\ref{eqn:survival1}) and 
$\psi_n \in \mbox{\rm Dom}(T)$, 
in the last equality. 
Note that above inequality holds for any $n \in {\bf N}$. 
Thus, in the limit $n \rightarrow \infty $, we obtain that 
%$\forall \psi \in {\cal H} $
\begin{equation}
\forall \psi \in {\cal H},~~~ 
\lim_{t \rightarrow \pm \infty } 
\left< \psi , e^{-itH} \psi \right> =0 . 
\label{eqn:decay}
\end{equation} 
This means that $H$ has no point spectrum. 
Because if $H$ has non-empty point spectrum, 
say $\lambda \in {\bf R}^1$, then there is a corresponding 
eigenvector $\psi_\lambda $, which satisfies  
$H \psi_\lambda = \lambda \psi_\lambda $. 
Obviously $\psi_\lambda $ does not satisfy 
the above condition (\ref{eqn:decay}). 
%
%and $\left< \psi_\lambda , e^{-itH} \psi_\lambda \right> 
%= e^{-it_\lambda } \| \psi_\lambda \| $. 
%
\hfill \qed

Moreover, it is seen that $H$ is absolutely continuous, 
under the same assumption as in Corollary \ref{co:Hpp0}. 
Its proof is, essentially, based on the theorem in \cite{re3-2}.
For later convenience, we here introduce 
the closed subspace of ${\cal H}$, 
with respect to a self-adjoint operator $H$ on ${\cal H}$, 
that is, 
%
%${\cal H}_{\rm pp} (H) := 
%\{ \psi \in {\cal H} ~|~ \| E( \cdot ) \psi \|^2 
%\mbox{ is pure point } \} $, 
%
${\cal H}_{\rm ac} (H)  := 
\{ \psi \in {\cal H} ~|~ \| E( \cdot ) \psi \|^2 
\mbox{ is absolutely continuous } \}$, 
%
%and 
%${\cal H}_{\rm sing} (H)  := 
%\{ \psi \in {\cal H} ~|~ \| E( \cdot ) \psi \|^2 
%\mbox{ is continuous singular} \}$ \cite{re-3}, 
%
where 
$\{ E(B) ~|~ B \in {\bf B}^1 \}$ is the spectral measure of $H$ 
\cite{re-3}.

\begin{th} \label{th:Hpp0} {\sl 
$:$  Let $T$ be a symmetric operator on ${\cal H}$, and 
$H$ be a self-adjoint operator on ${\cal H}$. 
If a pair of operators $T$ and $H$ 
satisfies the $T$-weak WR, then  
\begin{equation}
\| E(B) \psi \|^2 \leq \| T\psi \| \| \psi \| |B| 
\label{eqn:4.4.0.}
\end{equation} 
for all $\psi \in  \mbox{\rm Dom}(T) $ and all $B \in {\bf B}^1 $, 
where $|B|$ is the Lebesgue measure of $B$. 
In particular, $H$ is absolutely continuous. 
%, that is,   
%${\cal H}_{\rm sing} (H) = {\cal H}_{\rm pp} (H) =\{ 0\}$. 
}
\end{th}

{\sl Proof} :  Let us, first, derive the inequality 
that, $
\forall \epsilon >0,~~
\forall \lambda \in {\bf R}^1 ,~~
\mbox{ and } \forall \psi \in  \mbox{\rm Dom}(T) $,
\begin{equation}
\left| {\rm Im} \left< \psi ,  R( \lambda +i \epsilon ) 
\psi \right> \right| 
\leq  \pi \| T\psi \| \| \psi \| ,  
\label{eqn:4.4.1.}
\end{equation}
where $R(\lambda \pm i \epsilon) 
:= (H- (\lambda \pm i \epsilon))^{-1}$.  
It is seen that 
\begin{eqnarray}
i~ {\rm Im} 
\left< \psi , R( \lambda +i \epsilon ) \psi \right> 
& = & \frac{1}{2} \int_{{\bf R}^1 } 
      \left( \frac{1}{\lambda^{\prime} - \lambda -i \epsilon  }
      - \frac{1}{\lambda^{\prime} - \lambda +i \epsilon  }
      \right) d \left< \psi , E(\lambda^{\prime} ) \psi \right> 
      \nonumber \\
& = & \frac{i}{2} \int_{{\bf R}^1 } 
      \left[ 
      \int_0 ^{\infty } 
      ( e^{-it (\lambda^{\prime} - \lambda -i \epsilon ) } + 
      e^{it (\lambda^{\prime} - \lambda + i \epsilon ) } ) dt 
      \right] 
      d \left< \psi , E(\lambda^{\prime} ) \psi \right> 
      \nonumber \\
& = & i \int_0 ^{\infty } 
      e^{-\epsilon t } 
      \left< \psi , \cos t( H - \lambda ) \psi \right> dt 
      \nonumber \\
& = & \lim_{ \delta \downarrow 0 } 
      \int_{\delta } ^{\infty } 
      \frac{e^{-\epsilon t } }{t } 
      \left< \psi , [T, \sin t( H - \lambda ) ] \psi \right> dt,  
      \label{eqn:4.4.2.}
\end{eqnarray}
where, Fubini's theorem 
has been used in the third equality,  
and Eq.\ (\ref{eqn:survival2}) in the last. 
To evaluate Eq.\ (\ref{eqn:4.4.2.}), it is sufficient to 
see that 
\begin{equation}
    \lim_{ \delta \downarrow 0 } 
    \int_{\delta } ^{\infty } 
    \frac{e^{-\epsilon t } }{t } 
    \left< T \psi ,  \sin t( H - \lambda )  \psi \right> dt . 
    \label{eqn:4.4.3.}
\end{equation}
We define, here, a function $f(\epsilon , \lambda ) : 
(0, \infty ) \times {\bf R}^1 \rightarrow {\bf R}^1 $, 
as follows, 
\[
f(\epsilon , \lambda ) : = 
\int_0 ^{\infty } e^{-\epsilon t } 
\frac{ \sin t \lambda }{t} dt .
\]
$f(\epsilon , \lambda )$ is continuous on 
$(0, \infty ) \times {\bf R}^1 $, 
because of the fact that  
$|e^{-\epsilon t } \sin t \lambda /t | 
\leq e^{-\epsilon t } |\lambda |$ for any $t >0$, 
and of the use of the dominated convergence theorem. %\cite{re}.  
% e. g. the closed graph theorem 
Furthermore, since 
$\forall \epsilon >0 ,~ \forall \lambda \in {\bf R}^1$ 
$e^{-\epsilon t } \sin t \lambda $ is integrable on $[0, \infty )$, 
$f(\epsilon , \lambda )$ is differentiable 
with respect to any $\epsilon >0$, 
for each fixed $\lambda$. Thus, it is obtained, 
through the partial integrations,  that $\forall \lambda \neq 0$, 
\[ 
\partial_{\epsilon} f(\epsilon , \lambda ) 
= -\frac{1}{\lambda } \frac{1}{1+ \epsilon^2 / \lambda^2 } . 
\]
Note that $\forall \lambda \in {\bf R}^1 ,~ 
\lim_{\epsilon \rightarrow \infty } 
f(\epsilon , \lambda ) = 0 $, we obtain 
\begin{equation}
f(\epsilon , \lambda ) = \pm \frac{\pi}{2} 
-\frac{1}{\lambda } \int_{0} ^{\epsilon }
\frac{1}{1+ \tau^2 / \lambda^2 } d \tau ,   
\label{eqn:4.4.4.}
\end{equation}
where each $\pm$ corresponds to the sign of $\lambda $. 
From this expression, 
$f(\epsilon , \lambda ) $ is bounded, i. e. 
$| f(\epsilon , \lambda ) | \leq \pi /2 $. 
Eq.\ (\ref{eqn:4.4.3.}) is expressed by $f(\epsilon , \lambda ) $,  
\begin{eqnarray*}
    \lim_{ \delta \downarrow 0 } 
    \int_{\delta } ^{\infty } 
    \frac{e^{-\epsilon t } }{t } 
    \left< T \psi ,  \sin t( H - \lambda )  \psi \right> dt 
& = &   \lim_{ \delta \downarrow 0 } 
    \int_{\delta } ^{\infty } 
    \frac{e^{-\epsilon t } }{t } 
    \left[ 
    \int_{{\bf R}^1 } 
    \sin t( \lambda^{\prime } - \lambda )
    d \left< T \psi ,  E(  \lambda^{\prime } )  \psi \right> 
    \right] 
    dt \\ 
& = &   \lim_{ \delta \downarrow 0 } 
    \int_{{\bf R}^1 } 
    \left[ 
    \int_{\delta } ^{\infty } 
    e^{-\epsilon t } 
    \frac{
    \sin t( \lambda^{\prime } - \lambda )}{t } 
    dt
    \right] 
    d \left< T \psi ,  E(  \lambda^{\prime } )  \psi \right> \\ 
& = & 
    \int_{{\bf R}^1 } 
    f(\epsilon , \lambda^{\prime } - \lambda  ) 
    d \left< T \psi ,  E(  \lambda^{\prime } )  \psi \right> \\ 
& = & \left< T \psi , f(\epsilon , H -\lambda )   \psi \right> , 
%    \label{eqn:4.4.3.}
\end{eqnarray*}
where Fubini's theorem is used in the second equality, and 
the dominated convergence theorem is in the third. 
Substituting above relation into Eq.\ (\ref{eqn:4.4.2.}), 
\[ 
i~ {\rm Im} 
\left< \psi , R( \lambda +i \epsilon ) \psi \right> 
=  
\left< T \psi , f(\epsilon , H -\lambda )   \psi \right> - 
\left< f(\epsilon , H -\lambda )   \psi , T \psi \right>  . 
\]
Note that  $\| f(\epsilon , H -\lambda )  \| \leq \pi /2$, 
then Eq.\ (\ref{eqn:4.4.1.}) is obtained.  
Eq. (\ref{eqn:4.4.0.}) follows from Eq.\ (\ref{eqn:4.4.1.})  
through Stone's formula. 
By virtues of Eq.\ (\ref{eqn:4.4.0.}) and 
the denseness of 
$\mbox{\rm Dom}(T)$ in ${\cal H}$, 
%it is shown that, for every open interval $(a,b)$ 
%($-\infty < a < b < \infty $), 
%$E((a,b)) {\cal H} \subset {\cal H}_{\rm ac} (H) $ 
%\cite{cy}.  
%This implies that ${\cal H} = {\cal H}_{\rm ac} (H) $,  
%and therefore, 
it is seen that $H$ is absolutely continuous. 
\hfill \qed

\section{Absence of minimum-uncertainty states}
\label{sec:5}

\setcounter{df}{0}

When a pair of operators $T$ and $H$ 
satisfies the $T$-weak WR, 
the following uncertainty relation between them 
\begin{equation}
 \hspace{15mm}
 \left( \Delta T\right)_{\psi} \left( \Delta H\right)_{\psi} 
 \geq \frac{1}{2} , ~~~~~\forall \psi \in \mbox{\rm Dom}(TH) \cap 
\mbox{\rm Dom}(HT) ~~(\| \psi \| =1)
\label{eqn:uncertainty}
\end{equation}
is automatically derived, 
from the CCR between $T$ and $H$, the validity of which follows 
from Proposition\ \ref{pp:T-weakWRtoCCR} 
(a more detailed explanation will be given in the proof of Theorem\ 
\ref{th:uncertainty}). 
For operators $Q$ and $P$ in Sec.\ \ref{sec:3}, 
it is well known that there is a state 
$\psi \in \mbox{\rm Dom}(QP) \cap 
\mbox{\rm Dom}(PQ)$ ($\| \psi \| =1$), which minimizes 
the uncertainty, 
that is, a Gaussian packet. 
The following statement, on the contrary, shows 
that, under some additional conditions, there is 
no state $\psi \in \mbox{\rm Dom}(TH) \cap \mbox{\rm Dom}(HT)$ 
($\| \psi \| =1$) which satisfies the equality in Eq.\ 
(\ref{eqn:uncertainty}).

\begin{th} \label{th:uncertainty} {\sl 
: Let $T$ be a symmetric operator on ${\cal H}$, 
$H$ be a self-adjoint operator on ${\cal H}$, and 
these operators 
satisfy the $T$-weak WR. 
Then if $H$ is non negative 
and if the $T$-weak WR is analytically continued for all 
$t \in {\bf C}$ ( $\mbox{{\rm Im}~} t \leq 0 $),  
the equality in Eq.\ (\ref{eqn:uncertainty}) 
can never be satisfied by any 
$\psi \in \mbox{\rm Dom}(TH) \cap \mbox{\rm Dom}(HT)$ 
($\| \psi \| =1$).  }
\end{th}

In order to prove this theorem, 
let us first consider two lemmas.

{\sl 
\begin{lm} \label{lm:u1} {\sl : 
Let $T$ and $H$ be symmetric operators on ${\cal H}$, 
and they satisfy the CCR 
$TH - HT = i$, on a subspace of $\mbox{\rm Dom}(TH) \cap 
\mbox{\rm Dom}(HT)$, denoted by ${\cal D}$. 
Then neither eigenvector of $T$ nor that of $H$ belongs to 
${\cal D}$.}
\end{lm}}

{\sl Proof} : Assume that 
an eigenvector of $T$, $\psi _{\lambda } \neq 0$, 
belonging to an eigenvalue $\lambda$ exists in ${\cal D}$. 
Then 
$T\psi _{\lambda }=\lambda \psi_{\lambda }$, and thus 
we have 
$  \left< \psi _{\lambda } , (TH - HT) \psi _{\lambda }   \right>
% =  \left< T \psi _{\lambda } , H \psi _{\lambda }   \right> - 
%\left< H\psi _{\lambda } , T\psi _{\lambda }   \right> 
% =  \lambda \left< \psi _{\lambda } , H \psi _{\lambda }   \right> - 
% \lambda \left< \psi _{\lambda } , H\psi _{\lambda }   \right> 
  =  0$. 
On the other hand, the condition in 
this lemma requires that  
$ \left< \psi _{\lambda } 
, (TH-HT)\psi _{\lambda }   \right> =i\| \psi _{\lambda }  \| ^2 
\neq 0$. 
Thus the subspace ${\cal D}$ contains no 
eigenvector of $T$. 
%a contradiction comes out and $\psi _{\lambda } \notin 
%  \mbox{\rm Dom}(TH-HT)$ 
% should hold. 
The rest of the proof for $H$ can be done, as in 
the same way for $T$. ~~\hfill \qed

{\sl 
\begin{lm} \label{lm:u2} {\sl : 
Let $T$ and $H$ be symmetric operators on ${\cal H}$, 
and them satisfy the CCR 
$TH - HT = i$, on a subspace of $\mbox{\rm Dom}(TH) \cap 
\mbox{\rm Dom}(HT)$, denoted by ${\cal D}$. 
If a state $\eta \in {\cal D}$ ($\| \eta \| =1$) and 
a pair of complex numbers $a,b \in {\bf C}$,  
satisfying the following two equalities 
\begin{eqnarray}
& (T+aH+b)\eta =0 , &
 \label{eqn:u21} \\
& \left< T\eta , H\eta   \right> +\left< H\eta , T\eta   
\right> -2\left< T \right> _{\eta } \left< H \right> _{\eta } =0 , 
 & \label{eqn:u22} 
\end{eqnarray}
%\begin{equation}
% (T+aH+b)\eta =0 
% \label{eqn:u21} 
%\end{equation}
%and 
%\begin{equation}
% \left< T\eta , H\eta   \right> +\left< H\eta , T\eta   
%\right> -2\left< T \right> _{\eta } \left< H \right> _{\eta } =0 , 
% \label{eqn:u22} 
%\end{equation}
exist, then $ \mbox{{\rm Re}~} a =0$ and $ \mbox{{\rm Im}~} a > 0 $.} 
\end{lm}}

{\sl Proof} : Let $\eta$ be a state which satisfies 
the conditions in this lemma. Then we have that 
\[
        \left< T\eta , H\eta   \right> +\left< H\eta , T\eta   \right> 
       =  \left< \eta , (HT+i)\eta   \right> 
       +\left< H\eta , T\eta   \right>  
       = i -2a\| H\eta \|^2 -2b \left< \eta , H\eta 
         \right>  ~.
\]
%\begin{eqnarray}
%        \left< T\eta , H\eta   \right> +\left< H\eta , T\eta   \right> 
%       &= & \left< \eta , (HT+i)\eta   \right> 
%       +\left< H\eta , T\eta   \right>  
%       \nonumber \\
%       &= & i\| \eta \|^2 -2\left< H\eta , (aH+b)\eta   \right>  
%       \nonumber \\
%       &= & i -2a\| H\eta \|^2 -2b \left< \eta , H\eta 
%         \right>  ~.
%       \nonumber
%       \label{eqn:u23}
%\end{eqnarray}
%where we use the facts that 
%$TH - HT = i$ on ${\cal D}$, Eq.\ (\ref{eqn:u21}), 
%and $\| \eta \| =1$, in this turn. 
%It is also obtained, 
From Eq.\ (\ref{eqn:u21}), we also have that 
\[
        2 \left< T \right> _{\eta } \left< H \right> _{\eta }  
        = -2\left< \eta  , (aH+b)\eta   \right>  
       \left< \eta  , H\eta   \right>  
        = -2a\left< \eta  , H\eta    \right> ^2 -2b 
        \left< \eta  , H\eta  
         \right> .
         \]
%Substituting these Eqs.\ (\ref{eqn:u23}) and (\ref{eqn:u24}) 
%into Eq.\ (\ref{eqn:u22}), we have 
Therefore the condition Eq.\ (\ref{eqn:u22}) 
leads us to the relation 
\begin{equation}
 i -2a\| H\eta \|^2 =  -2a\left< \eta , H\eta   \right> ^2  . 
 \label{eqn:u25}
\end{equation}
Let us consider the real and 
imaginary parts of the above equality, separately. 
It follows, from the real part 
$ (\mbox{{\rm Re}~} a) \| H\eta \|^2 =  (\mbox{{\rm Re}~} a) 
\left< \eta , H\eta   \right> ^2 $, 
that $ \mbox{{\rm Re}~} a = 0 $. 
This is because if  $\mbox{{\rm Re}~} a  \neq 0$, then 
$\| H\eta \| ^2 -\left< \eta , H\eta   \right>^2 =0 $ 
and this means that $\eta$ is 
%
%$\| H\eta \| ^2 -\left< \eta , H\eta   \right>^2 
%=\| H\eta -\left< \eta , H\eta   \right> \eta \| ^2 ,~
%\| \eta \| =1$ holds to lead that $\eta \in 
%\mbox{\rm Dom}(TH) \cap \mbox{\rm Dom}(HT)$ must be 
an eigenvector of $H$, belonging to the eigenvalue 
$\left< \eta , H\eta   \right>$, 
in spite of the premise $\eta \in {\cal D}$ 
($\| \eta \| =1$). 
This is in contradiction to Lemma\ \ref{lm:u1}. 
It is also seen, 
from the imaginary part, 
$1-2 (\mbox{{\rm Im}~} a) \| H\eta \|^2 =  -2 
(\mbox{{\rm Im}~} a) \left< \eta , H\eta   \right>^2 $,  
that $ \mbox{{\rm Im}~} a > 0 $. 
%Because if $ \mbox{{\rm Im}~} a \leq 0 $ holds, then this derives 
%a following contradiction that  
%$1+2| \mbox{{\rm Im}~} a | \| H\eta \|^2 =  2 | \mbox{{\rm Im}~} 
%a | \left< \eta , H\eta   \right>^2 \leq 2 | \mbox{{\rm Im}~} 
%a| \| H\eta \|^2 $. 
~~\hfill \qed

{\sl Proof of Theorem\ \ref{th:uncertainty}} : Let $\psi \in 
\mbox{\rm Dom}(TH) \cap \mbox{\rm Dom}(HT)$ and $\| \psi \| =1$. 
Since the $T$-weak WR holds for $T$ and $H$, 
%from the condition in this theorem, 
the CCR in Eq.\ (\ref{eqn:1.2}) follows. 
%holds from Proposition\ 
%\ref{pp:T-weakWRtoCCR}.  
Then the uncertainty relation 
between $T$ and $H$, in Eq.\ (\ref{eqn:uncertainty}), 
is derived as 
%are derived 
%from the usual calculation 
%as, 
\begin{eqnarray*}
       \left( \Delta T \right) _{\psi}   
       \left( \Delta  H \right) _{\psi }  
       & = &  \| (T- \left< T \right> _{\psi})\psi \|  
       \| (H- \left< H \right> _{\psi})\psi \|   \\
       & \geq & \left| \left< (T- \left< T \right> _{\psi})\psi ,  
       (H- \left< H \right> _{\psi})\psi   \right> \right|  \\
       & \geq &  \left|  \mbox{{\rm Im}~} \left< (T- \left< T \right> _{\psi})\psi , (H- \left< H \right> _{\psi})\psi   \right> \right| \\
      & = &  \frac{1}{2}  \left| \left< T\psi 
      , H\psi   \right> - \left< H\psi , T\psi   \right> \right| 
      =  \frac{1}{2} ~~.
\end{eqnarray*}
In the second line, which is nothing but the Cauchy-Schwarz 
inequality, the equality holds if and only if 
there exists a complex number $\alpha \in {\bf C}$, 
satisfying  
\[ (T- \left< T \right> _{\psi})\psi +
\alpha (H- \left< H \right> _{\psi})\psi  =0 . \] 
In the third line, the equality holds if and only if 
\[ \mbox{{\rm Re}~} \left< 
(T- \left< T \right> _{\psi})\psi , (H- \left< H \right> _{\psi})\psi 
\right>  
= \left< T \psi ,  H\psi   \right> + \left< H\psi , T\psi   \right> -2 \left< T \right> _{\psi}  \left< H \right> _{\psi} =0 .\] 

\noindent 
In order to show that no $\psi \in \mbox{\rm Dom}(TH) 
\cap \mbox{\rm Dom}(HT)$ $(\| \psi \| =1)$ 
can satisfy the equality 
in the uncertainty relation between $T$ and $H$, 
in Eq.\ (\ref{eqn:uncertainty}), 
it is sufficient to 
see  that, 
above two conditions can not be satisfied simultaneously, 
for any $\psi \in \mbox{\rm Dom}(TH) 
\cap \mbox{\rm Dom}(HT)$ $(\| \psi \| =1)$ and 
for any $\alpha \in {\bf C}$. 
Observe that these conditions take just the same form as the two 
equalities (\ref{eqn:u21}) and (\ref{eqn:u22}) 
in Lemma\ \ref{lm:u2}.

Let us now assume that there exist such a state 
$\eta \in \mbox{\rm Dom}(TH) 
\cap \mbox{\rm Dom}(HT)$ ($\| \eta \| =1 $), and 
a pair of complex numbers 
$a,b \in {\bf C}$, that satisfy both of 
Eqs.\ (\ref{eqn:u21}) and (\ref{eqn:u22}), 
and derive a contradiction. 
Lemma\ \ref{lm:u2} implies that the parameter $a$ is pure 
imaginary and is expressed as $a=iq,~q > 0$, to lead 
$T\eta + iq H\eta + b\eta =0$. %for any $b \in {\bf C}$ and 
%any $\eta \in  \mbox{\rm Dom}([T,H]),~\| \eta \| =1$.
%
Then we must have that 
$- q\left< \eta , H\eta  \right> = \mbox{{\rm Im}~} b \leq 0$, 
because $\left< \eta , T\eta   \right> \in {\bf R}^1 $ and 
$ H \geq 0$, (see the conditions of Theorem\ \ref{th:uncertainty}). 
It is also noted that 
$e^{-itH}$ is bounded and 
$e^{-itH}H \subset H e^{-itH}$, 
for all $t \in {\bf C}$ 
($ \mbox{{\rm Im}~} t \leq 0 $). 
Then it follows that 
$Te^{-itH} \eta  =  e^{-itH} (T\eta + t\eta ) 
= ( - iq H - b + t )e^{-itH} \eta$. 
Since $ \mbox{{\rm Im}~} b \leq 0$ and 
$ \mbox{{\rm Im}~} t \leq 0$, we can put $t=b$ and 
obtain $Te^{-ibH} \eta = - iq He^{-ibH} \eta$. 
It is here noted that 
$e^{-ibH} \eta \neq 0$, because $e^{-ibH}$ is an injection 
and $\eta \neq 0$. 
%if $e^{-ibH} \eta = 0$ holds, 
%then $ \eta  =0$ must hold, against to 
%the condition $\| \eta \| =1$. 
This is seen from the following relations,  
\begin{eqnarray*}
        \| e^{-ib H } \eta \| ^2 
         & = & 
%\int_{{\bf R}^1 } |e^{-ib \lambda  }|^2 d\|
%E(\lambda ) \eta \| ^2 =
           \int_{[0,\infty ) } |e^{-ib \lambda  }|^2 d\|
E(\lambda ) \eta \| ^2 
           \geq  \int_{[0,N] } |e^{-ib \lambda  }|^2 d\|
E(\lambda ) \eta \| ^2 \\
         & \geq & \inf_{\lambda \in [0,N]}|e^{-ib \lambda
}|^2 ~\int_{[0,N] } d\| E(\lambda ) \eta \| ^2 
          =  e^{-2 \bigl| {\footnotesize \mbox{{\rm Im}~} } b 
          \bigr| N} 
~\int_{[0,N] } d\| E(\lambda ) \eta \| ^2 , 
\end{eqnarray*}
where $\{ E(B) ~|~ B \in {\bf B}^1 \}$ is the specral measure of $H$, 
$N$ is an arbitrary natural number, and  
we have used 
the fact that the spectrum of $H$ should be included in 
$[0, \infty )$, 
because $H \geq 0$. 
Providing that $e^{-ibH} \eta = 0$, one would obtain  
that $0= \lim_{N \rightarrow \infty } \displaystyle{ 
\int_{[0,N] } d\| E(\lambda ) \eta \| ^2 } = 
\| \eta \|^2 $, which contradicts $\| \eta \| =1 $. 
%
%Because of $e^{-2|\Im t| N} \neq 0$, 
%$\int_{[0,N] } d\| E(\lambda ) \eta \| ^2 =0$ must hold for any 
%$N\in {\bf N}$, and thus, as $N\rightarrow \infty $, 
%we obtain $\| \eta \| =0$ and this is contradict to 
%the condition $\| \eta \| =1$. 
%
%
By taking the inner products between $e^{-ibH} \eta$ and 
each side of $Te^{-ibH} \eta = - iq He^{-ibH} \eta$, 
we have 
\[ \left< e^{-ibH} \eta , Te^{-ibH} \eta   \right> = - iq \left< 
 e^{-ibH} \eta , He^{-ibH} \eta   \right> .\] 
Notice that 
%the left-hand side of this equality is a real number, 
%whereas the right-hand side is pure imaginary and therefore 
the both sides of this equality have to vanish. 
%
%Because if $\left<  e^{-ibH} \eta , He^{-ibH} \eta   
%\right> =0$ should hold (note $q>0$), it could be obtained that 
%$H e^{-ibH} \eta =0$ (note $e^{-ibH} \eta \neq 0$). 
%This is seen from the calculation as 
%
Since $q>0$ and $e^{-ibH} \eta \neq 0$, we have that 
$0=\left<  e^{-ibH} \eta , He^{-ibH} \eta 
  \right> = \left<  H^{1/2}e^{-ibH} \eta , H^{1/2} e^{-ibH} \eta 
  \right> = \| H^{1/2}e^{-ibH} \eta \| ^2 $, 
where $H^{1/2}$ is a self-adjoint operator, satisfying 
$H=H^{1/2} H^{1/2}$ and $H^{1/2} \geq 0$. 
Thus $He^{-ibH} \eta = 0$. 
It is also seen that $e^{-ibH} \eta \in 
\mbox{\rm Dom}(TH) \cap \mbox{\rm Dom}(HT)$, 
because $\eta \in 
\mbox{\rm Dom}(TH) \cap \mbox{\rm Dom}(HT)$.  
These facts are in contradiction to Lemma\ \ref{lm:u1}. 
%Thus $\left<  e^{-ibH} \eta , He^{-ibH} \eta   \right> \neq 0$ 
%must hold.
%
%
Therefore, Eqs.   
(\ref{eqn:u21}) and (\ref{eqn:u22}) in Lemma\ \ref{lm:u2} 
can not be satisfied simultaneously. 
This means that the equality, 
in the uncertainty relation between 
$T$ and $H$ in Eq.\ (\ref{eqn:uncertainty}),  never holds 
under the condition of Theorem\ \ref{th:uncertainty}. 
~~\hfill \qed

The question about minimum-uncertainty states %, 
%which resulted in Theorem\ \ref{th:uncertainty}, 
is motivated 
by the following result by Kobe 
\cite{ko},  
%\begin{equation}
\[
   \lim _{ n \rightarrow \infty  }
   \left( \Delta \widetilde{T_0} \right)_{F^{ -1} \phi_n } 
   \left( \Delta H_0 \right)_{F^{ -1} \phi_n } 
   = \frac{1}{2} ~~,
\]
%   \label{eqn:kobe}
%\end{equation}
where $\phi_n$ ($n \geq 2$) is defined as 
in Example\ \ref{ex:1}. 
Note however, that $\phi_n$ does not converge in 
$L^2 $-norm, as $n \rightarrow \infty$. 
This Kobe's result 
implies the absence of minimum-uncertainty states. 
This result was also derived by Wigner\ \cite{wi} and 
Baute et al\ \cite{ba} , in different ways from ours.
It should be  notified that 
the absence of minimum-uncertainty states
expresses a crucial difference between 
the Weyl relation and the $T$-weak WR.

%\newpage

\section{Construction of the time operators 
for general quantum systems}
\label{sec:6}

\setcounter{df}{0}

We first summarize the several results so far obtained about 
the time operator $T_0$ in Eq.\ (\ref{eqn:1.1}),  
or an its extension $\widetilde{T_0}$ 
in Eq.\ (\ref{eqn:tilde{T_0}}), for the 1DFPS. 
%, which are understood from the previous sections. 
%
\begin{ex} \label{ex:4} {\em : 
$\widetilde{T_0}$ satisfies the $\widetilde{T_0}$-weak WR 
with $H_0$, as is seen in Eq.\ (\ref{eqn:T_0weakWR}). 
Then we have the following properties about $\widetilde{T_0}$.  
%\begin{flushleft}
%\begin{tabular}{ll}

\noindent (i)\ \  
The inequality (\ref{eqn:survival1}) 
between $\left( \Delta \widetilde{T_0} \right)_{\psi }$ 
and the survival probability of $\psi$ holds 
for all $\psi \in \mbox{\rm Dom}(\widetilde{T_0} )$ 
(Theorem \ref{th:survival}). 

\noindent (ii) \  $\widetilde{T_0}$ has no point spectrum 
(Corollary\ \ref{co:noneigenvalue}). 

\noindent (iii) \ The inequality (\ref{eqn:4.4.0.}) 
holds for all $\psi \in \mbox{\rm Dom}(\widetilde{T_0} )$ 
and all $B \in {\bf B}^1 $ 
(Theorem \ref{th:Hpp0}).

\noindent (iV) \  
The uncertainty relation (\ref{eqn:uncertainty}) 
between $\widetilde{T_0}$ and $H_0$ holds on $\mbox{\rm Dom}(\widetilde{T_0} H_0) 
 \cap \mbox{\rm Dom}(H_0 \widetilde{T_0}) $ 
(Proposition\ \ref{pp:T-weakWRtoCCR}), although 
there exists no state in $\mbox{\rm Dom}(\widetilde{T_0} H_0) 
\cap \mbox{\rm Dom}(H_0 \widetilde{T_0}) $, 
which satisfies equality in 
the uncertainty relation between $\widetilde{T_0}$ and $H_0$ 
(Theorem\ \ref{th:uncertainty}). 
%\end{tabular}
%\end{flushleft}
}
\end{ex}
%
%
%These are easily verified. 
It is seen, from the above (i), 
that the Gaussian packet $F \phi_0$ in Example\ \ref{ex:1} 
is not included in $\mbox{\rm Dom}(\widetilde{T_0} )$. 
Because, for large $|t|$, the survival probability of 
the Gaussian packet for the 1DFPS decays with 
an inverse-power law $|t|^{-1}$, and this is in contradiction to 
the behavior of the survival probability predicted by 
the inequality\ (\ref{eqn:survival1}). 
It is, however, noticed that this kind of estimation 
about the domain $\mbox{\rm Dom}(\widetilde{T_0} )$ is valid for 
one dimensional case. 
Because, as the dimension becomes higher, the survival probability 
decays faster, in general than $|t|^{-2}$. 
We also obtain, from the inequality (\ref{eqn:survival1}) 
for $\widetilde{T_0}$ and $H_0$, that 
\begin{equation}
%\[
 2\sqrt{2}\left( \Delta \widetilde{T_0} \right) _{\psi} \geq \tau_h 
(\psi ), 
%\]
\label{eqn:5.1}
\end{equation}
where $\tau_h (\psi )$ is defined as 
$\tau_h (\psi ):= \sup \left\{ t \geq 0~\left|~ 
\left| \left< \psi , e^{-itH_0} \psi 
\right> \right| ^2 = 1/2    \right. \right\} $, 
$\forall \psi \in\mbox{\rm Dom}(\widetilde{T_0})$ ($\| \psi \| =1)$. 
This relation is important, to give the direct connection 
between $\left( \Delta \widetilde{T_0} \right) _{\psi}  $ and 
the measurable quantity $\tau_h (\psi )$, 
although we don't know whether $\widetilde{T_0}$ itself 
is an observable. 
Let us also consider the physical meaning of the above (iii). 
The following inequality is derived from Eq.\ 
(\ref{eqn:4.4.0.}), 
\[ 
\| E_{H_0} (B) \psi \|^2 \leq 
\left( \Delta \widetilde{T_0} \right)_{\psi} |B| \] 
for all $\psi \in\mbox{\rm Dom}(\widetilde{T_0})$ 
($\| \psi \| =1)$ and $B \in {\bf B}^1$,  because of 
the fact that the $\widetilde{T_0}$-weak WR is not changed, 
with replacing $\widetilde{T_0}$ by $\widetilde{T_0} - 
\left< \widetilde{T_0} \right>_{\psi }$. 
Note that $\| E_{H_0} (B) \psi \|^2$ is the probability which 
one finds a measured energy-value in the range $B$ 
for the fixed $\psi$. 
Suppose that $\left( \Delta \widetilde{T_0} \right)_{\psi}$ is 
small, then the probability 
$\| E_{H_0} ( B ) \psi \|^2$ 
should be uniformly small for all $B \in {\bf B}^1$.  
This concludes that the probability distribution 
$\| E_{H_0} ( \cdot ) \psi \|^2$ 
has  a broad deviation, for 
$\| E_{H_0} ( {\bf R}^1 ) \psi \|^2 =1 $. 
Hence $\left( \Delta \widetilde{H_0} \right)_{\psi}$ 
must be large and this result is consistent with  
the uncertainty relation. 

%
%With respect to the above (iv), the following relation, given 
%by Kobe \cite{ko}, 
%%one can see, from the direct calculation,  that 
%\begin{equation}
%   \lim _{ n \rightarrow \infty  }
%   \left( \Delta T_0 \right)_{\phi_n } 
%   \left( \Delta H_0 \right)_{\phi_n } 
%   = \frac{1}{2} ~~,
%   \label{eqn:kobe}
%\end{equation}
%where $\phi_n$ ($n \geq 2$) is defined as 
%in Example\ \ref{ex:1}, should be noted. 
%Therefore we have ${\displaystyle 
% \inf _{ 
%          \| \psi \| =1,\ 
%           \psi \in {\footnotesize \mbox{\rm Dom}}([T_0 ,H_0 ])  
%          }
%   \left( \Delta T_0 \right)_{\psi} \left( \Delta H_0 \right)_{\psi} 
%   = \frac{1}{2} }$. %, 
%although the minimum uncertainty state does not exist. 
%

In order to see the existence of the other quantum systems, 
than the 1DFPS, 
for which a time operator exists, 
let us recall the results obtained by Putnam, 
in  the theory of 
Schr{\"o}dinger operators 
\cite{pu-2}. 
According to this theorem, 
if a potential $V(x)$ is a real-valued 
measurable function on ${\bf R}^1$ satisfying 
 $ 0 \leq V(x) \leq \mbox{\rm const},~\mbox{\rm a.e.}$ 
and $V(x) \in L^1 ({\bf R}^1)$, 
%and furthermore 
%the Hamiltonian $H_1$ on $L^2 ({\bf R}^1)$ is defined as 
%$ H_1 := H_0 + V(x) $, 
then $H_0$ and $H_1 := H_0 + V(x)$ 
defined on $L^2 ({\bf R}^1)$ are absolutely continuous, 
and furthermore the wave operators 
$ U_{\pm} := \mbox{s-}\lim_{t \rightarrow \pm \infty } 
e^{itH_1} e^{-itH_0} $ 
exist and are unitary operators satisfying 
$H_1 = U_{\pm} H_0 U_{\pm} ^* $. 
For our purpose, we first define the operators 
$T_{1,\pm } := U_{\pm} \widetilde{T_0} U_{\pm} ^* $ on $L^2 ({\bf R}^1)$, 
where $U_{\pm}$ are the wave operators 
defined in this Putnam's theorem. 
Then $T_{1,\pm }$ are symmetric and satisfy 
the $T_{1,\pm }$-weak WR with $H_1$, i.e.  
%\begin{equation}
\[
T_{1,\pm } e^{-itH_1} = e^{-itH_1} 
\left( T_{1,\pm } +t \right) , 
\] 
%\phi,~~~
%\forall \phi \in \mbox{\rm Dom}(T_{1,\pm }) ,~ 
% \forall t \in {\bf C} ~(\mbox{\rm Im } t \leq 0) 
% \label{eqn:5.2}
%\end{equation}
which are nothing but 
the unitary transformations of Eq.\ (\ref{eqn:T_0weakWR}). 
These operators $T_{1,\pm}$ are 
the time operators we have sought for other quantum systems 
than the 1DFPS, 
and they satisfy 
all the properties, described in Example\ \ref{ex:4}, 
with this $H_1$. 
%must be also valid for $T_{1,\pm }$ and $H_1$. 
%
%
%

For a quantum system which allows bound states, 
we can also construct a time operator satisfying the $T$-weak WR 
with the Hamiltonian $H$, by restricting it to act 
on the set of scattering states. 
The latters are usually identified with 
the subspace ${\cal H}_{\rm ac}(H)$ 
of the Hilbert space ${\cal H}$ under consideration. 
Because, in this case, the wave operator (if exits) 
is not a unitary operator on ${\cal H}$ in general,  
that is, the range 
$\mbox{\rm Ran}(U_{\pm} )$ becomes a proper subspace of 
${\cal H}$. 
In fact, according to Kuroda \cite{ku}, 
if the potential $V(x)$ is a real-valued 
measurable function on ${\bf R}^1$ satisfying 
$V(x) \in L^1 ({\bf R}^n) \cap L^2 ({\bf R}^n ),~n\leq 3$, 
and 
the Hamiltonian $H_1$ on $L^2 ({\bf R}^1)$ is defined as 
$ H_1 := H_0 + V(x) $, 
then the wave operators $U_{\pm}$ exist and are complete, i.e. 
$\mbox{\rm Ran}(U_{\pm}) = L_{\rm ac} ^2 (H_1)$, 
where $L_{\rm ac} ^2 (H_1) $ is a subspace 
in $L^2 ({\bf R}^1 )$, similarly 
defined as ${\cal H}_{\rm ac} (H)$ 
just before Theorem\ \ref{th:Hpp0} . 
As in the same way for Putnam's theorem, 
by the use of the wave operators $U_{\pm}$ 
defined for $H_0$ and $H_1$ in this Kuroda's theorem, 
we can define the operators 
$T_{1,\pm } := U_{\pm} \widetilde{T_0} U_{\pm} ^* $ 
on $L_{\rm ac} ^2 (H_1)$ 
and their domains,  
$\mbox{\rm Dom}(T_{1,\pm }) := U_{\pm} \mbox{\rm Dom}(\widetilde{T_0} )$. 
Then they are symmetric operators on $L_{\rm ac} ^2 (H_1)$, 
and satisfy 
%\begin{equation} 
\[
T_{1,\pm } e^{-itH_{1,ac}} = e^{-itH_{1,ac}} 
\left( T_{1,\pm } +t \right) . 
\]
%\phi,~~~
%\forall \phi \in \mbox{\rm Dom}(T_{1,\pm }) ,~ 
% \forall t \in {\bf C} ~(\mbox{\rm Im } t \leq 0) 
% \label{eqn:5.3}
%\end{equation} 
$H_{1,ac}$ is 
defined as $H_{1,ac} := H_1 |_{L_{\rm ac} ^2 (H_1)} 
= U_{\pm} H_0 U_{\pm} ^*$, and is 
called the (spectrally) absolutely continuous part of 
$H_1$ \cite{ka}. 
%defined in Kato's theorem \cite{ka}. 
By the unitary 
equivalence, Example\ \ref{ex:4} is also valid for the pair 
$T_{1,\pm}$ and $H_{1,{\rm ac}}$. 
We can not, however, extend these $T_{1,\pm}$ to 
the densely defined symmetric operators on $L^2 ({\bf R}^1 )$, 
so that they satisfy the $T_{1,\pm}$-weak WR with 
$H_1$, when $H_1$ has a point spectrum, i.e. 
$L_{\rm pp} ^2 (H_1) \neq \{ 0\}$. 
This is because such an extension 
contradicts Corollary\ \ref{co:Hpp0} and 
Theorem\ \ref{th:Hpp0}.   

%\newpage

\section{Concluding Remarks} \label{sec:7}         

\setcounter{df}{0}   

Analyzing the  $T$-weak Weyl relation ($T$-weak WR) 
in Eq.\ (\ref{eqn:1.4}), and obtaining 
several statements about the time operator, 
%are obtained, and 
we have seen that 
the Aharonov-Bohm time operator $T_0$ in Eq.\ (\ref{eqn:1.1}) 
(or an its symmetric extension $\widetilde{T_0}$ 
in Eq.\ (\ref{eqn:tilde{T_0}}) ) 
is characterized 
by the $\widetilde{T_0}$-weak WR in Eq.\ (\ref{eqn:T_0weakWR}). 
We have, in  particular, 
recognized the fact that %found from these results that 
the time operator is deeply connected to the survival probability. 
In relation to this considerable connection, 
we would like, first,  to revisit 
the inequality\ (\ref{eqn:survival1}) in 
Theorem\ \ref{th:survival}, Theorem\ \ref{th:Hpp0}, 
and their implications.

The inequality\ (\ref{eqn:survival1})  is important to bring us 
a possibility of understanding the time operator 
from the two different points. 
The first point is related to 
the measurement of the survival probability. 
Since the inequality\ (\ref{eqn:5.1}) 
derived from the inequality\ (\ref{eqn:survival1}) 
gives the quantitative relation between 
the standard deviation of the time operator $\widetilde{T_0}$ and 
the maximum half-time of the survival probability 
in the 1DFPS, 
%in the one-dimensional free particle system,  
%for the same state respectively, 
we may associate the time operator 
in quantum systems, 
with both the real and theoretical measurements 
of the survival probability. 
Another point is related to the connection with 
the dynamics of quantum systems. 
%
%the dynamical evolution of states. 
%
In order to see this possibility, we may 
refer to 
Proposition\ \ref{pp:rapidly}, and 
Corollary\ \ref{co:Hpp0} which is 
one of the applications of the inequality\ (\ref{eqn:survival1}). 
%This statement is consistent with the observation 
%in Sec.\ \ref{sec:6}. 
These facts imply a possibility of associating 
the time operator (or its domain), with the scattering state  
and its dynamics, through the survival probability. 
%
%
%Here we regard the subspace of scattering states as 
%
%
%
%

As a remark on Theorem\ \ref{th:Hpp0}, 
the following  suggestion by Putnam  
should be recalled, 
%We should here note the following  suggestion by Putnam, 
that is, 
the existence of the absolutely continuous part of the Hamiltonian  
can be inferred from the behavior of specific observables 
\cite{pu-3}. 
He considered the following system, in which 
there is a 
self-adjoint operator $A_0$ satisfying  
$ A_t = A_0 + tI  $, 
$\forall t \in {\bf R}^1 $, 
where $A_t := e^{itH} A_0 e^{-itH}$ and $H$ is the Hamiltonian for 
this system. He showed that $H$ must be absolutely continuous 
(note that this is the case to which the last statement 
in Proposition\ \ref{pp:T-weakWRtoCCR} is applicable). 
%This result is also understandable from the fact that, 
%in this case, $A$ and $H$ satisfy the Weyl relation 
%because of the same reason as in the comment on Proposition\ 
%\ref{pp:T-weakWRtoCCR}. 
%because of the discussion in Sec.\ \ref{sec:2}. 
The essence of its proof, that is, 
the uniqueness of the 
spectral resolution of $A_0$, 
also implies that if $A_0$ is maximally symmetric  
(not necessarily self-adjoint), 
$H$ must be absolutely continuous. 
Because $A_0$ is uniquely represented by 
the generalized resolution of identity \cite{ak2}. 
In this context, Theorem\ \ref{th:Hpp0} is a generalization 
of the above statement by Putnam, 
to non-maximally symmetric operators.  
%although it is not clear whether 
%the singular continuous part of $H$ does not exist. 

The proof of the absolute continuity of $H$, which satisfies 
the $T$-weak Weyl relation with $T$, depends on 
Eq.\ (\ref{eqn:4.4.0.}).
It is similar to the following inequality which 
was derived by Putnam \cite{pu-4} 
and Kato \cite{ka2}, on the study of the commutator of the form, 
$[H, iA] =C$, where $H, A$ and $C$ are bounded self-adjoint 
operators, and $C \geq 0$, 
\begin{equation}
\| C^{1/2} E_H (B) \psi \|^2 \leq \| A \|   \| \psi \|^2 |B|, 
~~~~~~\forall \psi \in {\cal H} ~~\forall B \in {\bf B}^1 . 
\label{eqn:7.0.0.}
\end{equation}
It is seen that Eq. (\ref{eqn:4.4.0.}) 
is an unbounded case of theirs, 
and  is not trivial. 
Because one can not replace directly $C$ with the identity, 
for both bounded $H$ and $A$. 
They showed, through Eq.\ (\ref{eqn:7.0.0.}), 
that $H$ is absolutely continuous, 
provided that $\mbox{\rm Ker}(C) =\{ 0 \} $, i.e. 
$\overline{\mbox{\rm Ran}(C)} ={\cal H}$. 
This statement is sometimes valid when the above $H$ and $A$ 
are unbounded, 
however, its proof  originates in 
the more fundamental notion, $T$-smoothness \cite{ka2}, rather than  
the inequality like the above one, 
and the appropriate technique. 
Lavine  applied it to Schr\"{o}dinger operators  
in an especially good manner \cite{la}, \cite{la2}, \cite{re3}. 
It should be noted that 
his work is closely related to the our problem.  
He found self-adjoint operators $A$ which satisfy 
$[H, iA] =C$, with a certain class of Schr\"{o}dinger operators 
$H$ and positive bounded $C$. 
In the case of the absolute continuity of the free Hamiltonian 
$H_0$, 
%our approach is possible to be reformulated,  
%by following the Lavine's. 
we can choose the self-adjoint operator $A:= (f(P)Q + Qf(P))/2$, 
where $f(P):=P (P^2 + \delta^2 )^{-1}$ and $C:=Pf(P)$. 
When $\delta >0$, $A$ is self-adjoint, because $\mbox{\rm Ker} 
(A^* \mp i) =\{ 0 \}$. 
And also $C$ is bounded and $C>0$.  
They satisfy $[H_0 , iA] =C$ and one can consider the case that 
a parameter $\delta$ approaches to $0$, which is just $A=T_0$. 
Hence this scheme has the advance in the approach to  
the absolute continuity of $H_0$ and its connection with 
the Aharonov-Bohm time operator $T_0$.

%\newpage

\section*{Acknowledgements} 
\setcounter{sn}{1}
\renewcommand{\thesn}{%
          \Alph{sn}}

The author would like to thank Professor I.\ Ohba 
and Professor H.\ Nakazato 
for useful and helpful comments 
and the referee for valuable comments. 
He would also like
to thank Drs. M.\ Hayasi,\  S.\ Osawa and K.\ Nakazawa 
for useful and helpful discussions.


\begin{thebibliography}{99}


\bibitem{ko} D.~H.~Kobe and V.~C.~Aguilera-Navarro, \newblock ``Derivation of 
the energy-time uncertainty relation,'' \newblock Phys.\ Rev.\ A50,\ 933-938\ 
(1994).  

\bibitem{aha} Y.~Aharonov and D.~Bohm, \newblock ``Time in the Quantum Theory 
and the Uncertainty Relation for Time and Energy,'' \newblock Phys.\ Rev.\ 
122,\ 1649-1658\ (1961).




\bibitem{pa} W.~Pauli, \newblock {\em Die allgemeinen Prinzipien 
der Wellenmechanik}, \newblock edited by S.~Fl{\"u}gge, 
\newblock Encyclopedia of Physics 
(Springer-Verlag,\ Berlin,\ 1958),\ Vol.\ V,\ pt.\ 1,\ pp.1-168. 
\newblock See footnote 1 on p.\ 60. 



\bibitem{eg} I.~L.~Egusquiza and J.~G.~Muga, \newblock 
``Free-motion time-of-arrival operator and probability distribution,'' 
\newblock Phys.\ Rev.\ A61,\ 012104\ (9 pages)\ (1999). 


\bibitem{ak} N.~I.~Akhiezer and I.~M.~Glazman, \newblock 
{\em Theory of Linear 
Operators in Hilbert Space},\ Vol.I,\ \newblock 
(Pitman Advanced Publishing Program,\ Boston,\ 1981),\ 
Chap.\ 4,\ Sec.\ 55.
% page 3





\bibitem{ga} E.~Galapon, \newblock ``The Consistency Of a Bounded, 
Self-Adjoint Time Operator Canonically Conjugate to a Hamiltonian 
with Non-empty Point spectrum,'' \newblock preprint,\ quant-ph/9908033. 


\bibitem{pu} C.~R.~Putnam, \newblock {\em Commutation Properties of Hilbert 
Space Operators and Related Topics}, \newblock 
(Springer-Verlag,\ Berlin,\  1967).




\bibitem{ka2} T.~Kato, \newblock ``Smooth operators and commutators,'' 
\newblock Stud.\ Math.\ 31, 135-148\ (1972).



\bibitem{la} R.~Lavine, \newblock ``Absolute continuity of 
Hamiltonian operators with repulsive potentials,'' \newblock 
Proc.\ Am.\ Math.\ Soc.\ 22,\ 55-60\ (1969).




\bibitem{la2} R.~Lavine, \newblock ``Commutators and scattering 
theory, I.\ Repulsive interactions,'' \newblock 
Comm.\ Math.\ Phys.\ 20,\ 301-323\ (1971).







\bibitem{re3} M.~Reed and B.~Simon, \newblock {\em Methods of Modern 
Mathematical Physics},\ Vol.III: {\em Scattering Theory} 
\newblock (Academic Press,\ New York,\ 1979 ),\ 
Chap.\ XIII,\ Sec.\ 6 and 7.





\bibitem{ka} T.~Kato, \newblock {\em Perturbation theory for linear 
operators}, \newblock 
(Springer-Verlag,\ Berlin,\  1966),\ 
Chap.\ X,\ Sec.\ 1.
% page 7






\bibitem{re} M.~Reed and B.~Simon, \newblock {\em Methods of Modern Mathematical Physics},\ Vol.I: {\em Functional Analysis} \newblock 
(Academic Press,\ New York,\ 1972 ),\  Chap.\ VIII,\ Sec.\ 5. 











\bibitem{re-2} Reference\ \cite{re},\ 
%\cite{re},\ 
Corollary of Theorem\ VIII.\ 14. 










\bibitem{ak-2} Reference\ \cite{ak},\ 
%\cite{ak},\  
%N.~I.~Akhiezer and I.~M.~Glazman, \newblock 
%{\em Theory of Linear 
%Operators in Hilbert Space},\ Vol.I,\ \newblock 
%(Pitman Advanced Publishing Program,\ Boston,\ 1981),\ 
Chap.\ 4,\ Sec.\ 46. 
% page 4




\bibitem{re3-3} Reference\ \cite{re3},\  
%M.~Reed and B.~Simon, \newblock {\em Methods of Modern 
%Mathematical Physics},\ Vol.III: {\em Scattering Theory} 
%\newblock (Academic Press,\ New York,\ 1979 ),\ 
Chap.\ XI,\ Sec.\ 3,\ Lemma\ 2.



\bibitem{bh} K.~Bhattacharyya, \newblock ``Quantum decay and the Mandelstam-Tamm time-energy inequality,'' \newblock J.\ Phys.\ A.\ 16,\ 2993-2996\ (1983).





\bibitem{re3-2} Reference\ \cite{re3},\ 
%Chap.\ XIII,\ Sec.\ 6,\ 
Theorem\ XIII.\ 20. 






\bibitem{re-3} Reference\ \cite{re},\ Chap.\ VII,\ Sec.\ 2. 







\bibitem{wi} E.~P.~Wigner, \newblock in {\em Aspects of Quantum 
Theory}, edited by\ A.~Salam\ and E.~P.~Wigner 
(Cambridge University Press,\ London\ 1972),\ 237-247.









\bibitem{ba} A.~D.~Baute,\ R.~Sala ~Mayato,\ J.~P.~Palao,\ 
J.~G.~Muga and I.~L.~Egusquiza, \newblock 
``Time-of-arrival distribution for arbitrary potentials 
and Wigner's time-energy uncertainty relation,'' \newblock 
Phys.\ Rev.\ A\ 61,\ 022118 (5 pages) (2000).









\bibitem{pu-2} Reference\ \cite{pu},\  
%\cite{pu},\ 
%Chap.\ V,\ Sec.\ 5.16,\ 
Theorem\ 5.16.2. 
% page 21


\bibitem{ku} S.~T.~Kuroda, \newblock ``On the existence and the unitary property of the scattering operator,'' \newblock Nuovo Cimento\ 12,\ 431-454\ (1959),\ 
Theorem\ 3.1;\ Theorem\ 5.1. 




\bibitem{pu-3} Reference\ \cite{pu},\  
%\cite{pu},\ 
Chap.\ II,\ Sec.\ 2.14.
% page 23





\bibitem{ak2} N.~I.~Akhiezer and I.~M.~Glazman, \newblock 
{\em Theory of Linear 
Operators in Hilbert Space},\ Vol.II,\ \newblock 
(Pitman Advanced Publishing Program,\ Boston,\ 1981),\ 
%Chap.\ 8. 
Chap.\ 9,\ Sec.\ 112.



\bibitem{pu-4} Reference\ \cite{pu},\  
%Chap.\ II,\ Sec.\ 2.13,\ 
Theorem\ 2.2.4.




%\bibitem{ak2-2} Reference\ \cite{ak2},\ Chap.\ 9,\ Sec.\ 112.



%\bibitem{pu-4} Reference\ \cite{pu},\  
%Chap.\ II,\ Sec.\ 2.13,\ 
%Theorem\ 2.13.2.




\end{thebibliography}
\end{document}